\documentclass[twocolumn]{aastex631}

\usepackage{IEEEtrantools} 
\usepackage{amsmath}

\newcolumntype{L}[1]{>{\raggedright\arraybackslash}p{#1}}
\newcolumntype{C}[1]{>{\centering\arraybackslash}p{#1}}
\newcolumntype{R}[1]{>{\raggedleft\arraybackslash}p{#1}}

\newcommand{\msun}{\ensuremath{M_{\odot}}}
\newcommand{\lsun}{\ensuremath{L_{\odot}}}
\newcommand{\rsun}{\ensuremath{R_{\odot}}}

\newcommand{\numax}{\ensuremath{\nu_{\textrm{max}}}}
\newcommand{\dnu}{\ensuremath{\Delta\nu}}

\newcommand{\teff}{\ensuremath{T_{\textrm{eff}}\:}}
\newcommand{\logg}{\mbox{$\log g$}}
\newcommand{\fbol}{\mbox{$f_{\rm{bol}}$}}

\newcommand{\muHz}{\mbox{$\mu$Hz}}
\newcommand{\feh}{\mbox{$\rm{[Fe/H]}$}}

\newcommand{\rhostar}{\mbox{$\rho_{\star}$}}
\newcommand{\rstar}{\mbox{$R_{\star}$}}

\newcommand{\mstar}{\mbox{$M_{\star}$}}

\providecommand{\mj}{\ensuremath{\,M_{\rm J}}}

\newcommand{\re}{\mbox{$R_{\ensuremath{\oplus}}$}}
\newcommand{\me}{\mbox{$M_{\ensuremath{\oplus}}$}}

\newcommand{\gpav}{\mbox{$\gamma$\,Pav}}
\newcommand{\ztuc}{\mbox{$\zeta$\,Tuc}}
\newcommand{\pmen}{\mbox{$\pi$\,Men}}

\newcommand{\kep}{\mbox{\textit{Kepler}}}
\newcommand{\tess}{{\it TESS}}
\newcommand{\gaia}{\mbox{\textit{Gaia}}}

\usepackage{xcolor}

\newcommand{\numaxgpav}{\mbox{$2693 \pm 95$}}
\newcommand{\dnugpav}{\mbox{$119.9 \pm 1.0$}}
\newcommand{\teffgpav}{\mbox{$6168 \pm 130$}}
\newcommand{\fehgpav}{\mbox{$-0.66 \pm 0.09$}}
\newcommand{\plxgpav}{\mbox{$108.01 \pm 0.11$}}
\newcommand{\lumgpav}{\mbox{$1.461 \pm 0.049$}}
\newcommand{\radgpav}{\mbox{$1.057\pm0.008 \rm{(ran)} \pm 0.009 \rm{(sys)}$}}
\newcommand{\massgpav}{\mbox{$0.934\pm0.021 \rm{(ran)} \pm 0.026 \rm{(sys)}$}}
\newcommand{\agegpav}{\mbox{$5.9\pm0.6 \rm{(ran)} \pm 1.0 \rm{(sys)}$}}
\newcommand{\logggpav}{\mbox{$4.359\pm0.003 \rm{(ran)} \pm 0.007 \rm{(sys)}$}}
\newcommand{\dengpav}{\mbox{$1.115\pm0.004 \rm{(ran)} \pm 0.014 \rm{(sys)}$}}
\newcommand{\fbolgpav}{\mbox{$5.47\pm0.18$}}

\newcommand{\numaxztuc}{\mbox{$2660 \pm 99$}}
\newcommand{\dnuztuc}{\mbox{$125.9 \pm 0.8$}}
\newcommand{\teffztuc}{\mbox{$5924 \pm 130$}}
\newcommand{\fehztuc}{\mbox{$-0.21 \pm 0.09$}}
\newcommand{\plxztuc}{\mbox{$116.18 \pm 0.13$}}
\newcommand{\lumztuc}{\mbox{$1.232 \pm 0.039$}}
\newcommand{\radztuc}{\mbox{$1.044\pm0.008 \rm{(ran)} \pm 0.006 \rm{(sys)}$}}
\newcommand{\massztuc}{\mbox{$0.985\pm0.020 \rm{(ran)} \pm 0.018 \rm{(sys)}$}}
\newcommand{\ageztuc}{\mbox{$5.3\pm0.5 \rm{(ran)} \pm 0.2 \rm{(sys)}$}}
\newcommand{\loggztuc}{\mbox{$4.394\pm0.003 \rm{(ran)} \pm 0.006 \rm{(sys)}$}}
\newcommand{\denztuc}{\mbox{$1.222\pm0.004 \rm{(ran)} \pm 0.017 \rm{(sys)}$}}
\newcommand{\fbolztuc}{\mbox{$5.34\pm0.17$}}

\newcommand{\numaxpmen}{\mbox{$2599 \pm 69$}}
\newcommand{\dnupmen}{\mbox{$116.7 \pm 1.1$}}
\newcommand{\teffpmen}{\mbox{$5980 \pm 130$}}
\newcommand{\fehpmen}{\mbox{$0.07 \pm 0.07$}}
\newcommand{\plxpmen}{\mbox{$54.683 \pm 0.035$}}
\newcommand{\lumpmen}{\mbox{$1.469 \pm 0.061$}}
\newcommand{\radpmen}{\mbox{$1.136\pm0.009 \rm{(ran)} \pm 0.006 \rm{(sys)}$}}
\newcommand{\masspmen}{\mbox{$1.091\pm0.026 \rm{(ran)} \pm 0.016 \rm{(sys)}$}}
\newcommand{\agepmen}{\mbox{$3.8\pm0.7 \rm{(ran)} \pm 0.4 \rm{(sys)}$}}
\newcommand{\loggpmen}{\mbox{$4.365\pm0.004 \rm{(ran)} \pm 0.006 \rm{(sys)}$}}
\newcommand{\denpmen}{\mbox{$1.050\pm0.003 \rm{(ran)} \pm 0.013 \rm{(sys)}$}}
\newcommand{\fbolpmen}{\mbox{$1.409\pm0.058$}}

\shorttitle{TESS 20-SECOND CADENCE DATA}  
\shortauthors{Huber et al.}

\begin{document}

\title{A 20-Second Cadence View of Solar-Type Stars and Their Planets with TESS: \\ Asteroseismology of Solar Analogs and a Re-characterization of $\pi$\,Men\,c}

\suppressAffiliations

\correspondingauthor{Daniel Huber}
\email{huberd@hawaii.edu}

\author[0000-0001-8832-4488]{Daniel Huber}
\affiliation{Institute for Astronomy, University of Hawai`i, 2680 Woodlawn Drive, Honolulu, HI 96822, USA}

\author[0000-0002-6980-3392]{Timothy R. White}
\affiliation{Sydney Institute for Astronomy (SIfA), School of Physics, University of Sydney, NSW 2006, Australia}
\affiliation{Stellar Astrophysics Centre (SAC), Department of Physics and Astronomy, Aarhus University, Ny Munkegade 120, DK-8000 Aarhus C, Denmark}

\author[0000-0003-4034-0416]{Travis S.\ Metcalfe}
\affiliation{White Dwarf Research Corporation, 9020 Brumm Trail, Golden, CO 80403, USA}

\author[0000-0003-1125-2564]{Ashley Chontos}
\affiliation{Institute for Astronomy, University of Hawai`i, 2680 Woodlawn Drive, Honolulu, HI 96822, USA}
\affiliation{NSF Graduate Research Fellow}

\author[0000-0002-9113-7162]{Michael M.\ Fausnaugh} %
\affiliation{Department of Physics, and Kavli Institute for Astrophysics and Space Research, Massachusetts Institute of Technology, 77 Massachusetts Ave., Cambridge, MA 02139, USA}

\author[0000-0001-7457-5120]{Cynthia S.\ K.\ Ho}
\author{Vincent Van Eylen}
\affiliation{Mullard Space Science Laboratory, University College London, Holmbury St Mary, Dorking, Surrey RH5 6NT, UK}


\author[0000-0002-4773-1017]{Warrick H.\ Ball} %
\affiliation{School of Physics and Astronomy, University of Birmingham, Birmingham B15 2TT, UK}
\affiliation{Stellar Astrophysics Centre (SAC), Department of Physics and Astronomy, Aarhus University, Ny Munkegade 120, DK-8000 Aarhus C, Denmark}

\author[0000-0002-6163-3472 ]{Sarbani Basu} %
\affiliation{Department of Astronomy, Yale University, P.O. Box 208101, New Haven, CT 06520-8101, USA}

\author[0000-0001-5222-4661]{Timothy R.\ Bedding} %
\affiliation{Sydney Institute for Astronomy (SIfA), School of Physics, University of Sydney, NSW 2006, Australia}
\affiliation{Stellar Astrophysics Centre (SAC), Department of Physics and Astronomy, Aarhus University, Ny Munkegade 120, DK-8000 Aarhus C, Denmark}

\author{Othman Benomar} %
\affiliation{Center for Space Science, New York University Abu Dhabi, UAE}
\affiliation{Division of Solar and Plasma Astrophysics, NAOJ, Mitaka, Tokyo, Japan}

\author{Diego Bossini}
\affiliation{Instituto de Astrof\'isica e Ci\^encias do Espa\c{c}o, Universidade do Porto, CAUP, Rua das Estrelas, 4150-762 Porto, Portugal}

\author{Sylvain Breton}
\affiliation{AIM, CEA, CNRS, Universit\'e Paris-Saclay, Universit\'e Paris Diderot, Sorbonne Paris Cit\'e, F-91191 Gif-sur-Yvette, France}

\author[0000-0002-1988-143X]{Derek L. Buzasi} %
\affiliation{Dept. of Chemistry \& Physics, Florida Gulf Coast University, 10501 FGCU Blvd. S., Fort Myers, FL 33965 USA}

\author[0000-0002-4588-5389]{Tiago L. Campante} %
\affiliation{Instituto de Astrof\'isica e Ci\^encias do Espa\c{c}o, Universidade do Porto, CAUP, Rua das Estrelas, 4150-762 Porto, Portugal}
\affiliation{Departamento de F\'{\i}sica e Astronomia, Faculdade de Ci\^{e}ncias da Universidade do Porto, Rua do Campo Alegre, s/n, 4169-007 Porto, Portugal}

\author[0000-0002-5714-8618]{William J. Chaplin} %
\affiliation{School of Physics and Astronomy, University of Birmingham, Birmingham B15 2TT, UK}
\affiliation{Stellar Astrophysics Centre (SAC), Department of Physics and Astronomy, Aarhus University, Ny Munkegade 120, DK-8000 Aarhus C, Denmark}

\author[0000-0001-5137-0966]{J\o rgen Christensen-Dalsgaard} %
\affiliation{Stellar Astrophysics Centre (SAC), Department of Physics and Astronomy, Aarhus University, Ny Munkegade 120, DK-8000 Aarhus C, Denmark}

\author[0000-0001-8237-7343]{Margarida S. Cunha} %
\affiliation{Instituto de Astrof\'isica e Ci\^encias do Espa\c{c}o, Universidade do Porto, CAUP, Rua das Estrelas, 4150-762 Porto, Portugal}
\affiliation{Departamento de F\'{\i}sica e Astronomia, Faculdade de Ci\^{e}ncias da Universidade do Porto, Rua do Campo Alegre, s/n, 4169-007 Porto, Portugal}

\author[0000-0001-6774-3587]{Morgan Deal} %
\affiliation{Instituto de Astrof\'isica e Ci\^encias do Espa\c{c}o, Universidade do Porto, CAUP, Rua das Estrelas, 4150-762 Porto, Portugal}
\affiliation{Departamento de F\'{\i}sica e Astronomia, Faculdade de Ci\^{e}ncias da Universidade do Porto, Rua do Campo Alegre, s/n, 4169-007 Porto, Portugal}

\author[0000-0002-8854-3776]{Rafael ~A.~Garc\'\i a} %
\affiliation{AIM, CEA, CNRS, Universit\'e Paris-Saclay, Universit\'e Paris Diderot, Sorbonne Paris Cit\'e, F-91191 Gif-sur-Yvette, France}

\author{Antonio Garc\'ia Mu\~noz} %
\affiliation{AIM, CEA, CNRS, Universit\'e Paris-Saclay, Universit\'e Paris Diderot, Sorbonne Paris Cit\'e, F-91191 Gif-sur-Yvette, France}

\author[0000-0002-0833-7084]{Charlotte Gehan} %
\affiliation{Instituto de Astrof\'isica e Ci\^encias do Espa\c{c}o, Universidade do Porto, CAUP, Rua das Estrelas, 4150-762 Porto, Portugal}
\affiliation{Max-Planck-Institut f\"ur Sonnensystemforschung, Justus-von-Liebig-Weg 3, 37077 G\"ottingen, Germany}

\author{Luc\'\i a Gonz\' alez-Cuesta}
\affiliation{Instituto de Astrof\'isica de Canarias (IAC), 38205 La Laguna, Tenerife, Spain}
\affiliation{Universidad de La Laguna (ULL), Departamento de Astrof\'isica, E-38206 La Laguna, Tenerife, Spain}

\author[0000-0002-7614-1665]{Chen Jiang} %
\affiliation{Max-Planck-Institut f\"ur Sonnensystemforschung, Justus-von-Liebig-Weg 3, 37077 G\"ottingen, Germany}

\author{Cenk Kayhan} %
\affiliation{Department of Astronomy and Space Sciences, Science Faculty, Erciyes University, 38030 Melikgazi, Kayseri, Turkey}

\author{Hans Kjeldsen} %
\affiliation{Stellar Astrophysics Centre (SAC), Department of Physics and Astronomy, Aarhus University, Ny Munkegade 120, DK-8000 Aarhus C, Denmark}
\affiliation{Institute of Theoretical Physics and Astronomy, Vilnius University, Sauletekio av. 3, 10257 Vilnius, Lithuania}

\author[0000-0002-8661-2571]{Mia S.\ Lundkvist} %
\affiliation{Stellar Astrophysics Centre (SAC), Department of Physics and Astronomy, Aarhus University, Ny Munkegade 120, DK-8000 Aarhus C, Denmark}

\author{St\'ephane Mathis} %
\affiliation{AIM, CEA, CNRS, Universit\'e Paris-Saclay, Universit\'e Paris Diderot, Sorbonne Paris Cit\'e, F-91191 Gif-sur-Yvette, France}

\author[0000-0002-0129-0316]{Savita Mathur} %
\affiliation{Instituto de Astrof\'isica de Canarias (IAC), 38205 La Laguna, Tenerife, Spain}
\affiliation{Universidad de La Laguna (ULL), Departamento de Astrof\'isica, E-38206 La Laguna, Tenerife, Spain}

\author[0000-0003-0513-8116]{M\'ario J. P. F. G. Monteiro} %
\affiliation{Instituto de Astrof\'isica e Ci\^encias do Espa\c{c}o, Universidade do Porto, CAUP, Rua das Estrelas, 4150-762 Porto, Portugal}
\affiliation{Departamento de F\'{\i}sica e Astronomia, Faculdade de Ci\^{e}ncias da Universidade do Porto, Rua do Campo Alegre, s/n, 4169-007 Porto, Portugal}

\author[0000-0002-4647-2068]{Benard Nsamba} %
\affiliation{Max-Planck-Institut f\"{u}r Astrophysik, Karl-Schwarzschild-Str. 1, D-85748 Garching, Germany}
\affiliation{Instituto de Astrof\'{\i}sica e Ci\^{e}ncias do Espa\c{c}o, Universidade do Porto,  Rua das Estrelas, PT4150-762 Porto, Portugal}
\affiliation{Kyambogo University, P.O. Box 1, Kyambogo Hill, Kampala - Uganda}

\author[0000-0001-7664-648X]{Jia Mian Joel Ong} %
\affiliation{Department of Astronomy, Yale University, P.O. Box 208101, New Haven, CT 06520-8101, USA}

\author{Erika Pak\v{s}tien\.{e}} %
\affiliation{Institute of Theoretical Physics and Astronomy, Vilnius University, Sauletekio av. 3, 10257 Vilnius, Lithuania}

\author[0000-0001-6359-2769]{Aldo M. Serenelli} %
\affiliation{Institute of Space Sciences (ICE, CSIC) Campus UAB, Carrer de Can Magrans, s/n, E-08193, Barcelona, Spain}
\affiliation{Institut d’Estudis Espacials de Catalunya (IEEC), C/Gran Capita, 2-4, E-08034, Barcelona, Spain}

\author[0000-0002-6137-903X]{Victor Silva Aguirre} %
\affiliation{Stellar Astrophysics Centre (SAC), Department of Physics and Astronomy, Aarhus University, Ny Munkegade 120, DK-8000 Aarhus C, Denmark}

\author[0000-0002-3481-9052]{Keivan G. Stassun} %
\affiliation{Vanderbilt University, Department of Physics \& Astronomy, 6301 Stevenson Center Ln., Nashville, TN 37235, USA}
\affiliation{Vanderbilt Initiative in Data-intensive Astrophysics (VIDA), 6301 Stevenson Center Lane, Nashville, TN 37235, USA}

\author{Dennis Stello} %
\affiliation{School of Physics, The University of New South Wales, Sydney NSW 2052, Australia}
\affiliation{Sydney Institute for Astronomy (SIfA), School of Physics, University of Sydney, NSW 2006, Australia}
\affiliation{Stellar Astrophysics Centre (SAC), Department of Physics and Astronomy, Aarhus University, Ny Munkegade 120, DK-8000 Aarhus C, Denmark}

\author{Sissel Norgaard Stilling} %
\affiliation{Stellar Astrophysics Centre (SAC), Department of Physics and Astronomy, Aarhus University, Ny Munkegade 120, DK-8000 Aarhus C, Denmark}

\author{Mark Lykke Winther} %
\affiliation{Stellar Astrophysics Centre (SAC), Department of Physics and Astronomy, Aarhus University, Ny Munkegade 120, DK-8000 Aarhus C, Denmark}

\author[0000-0001-6832-4325]{Tao Wu} %
\affiliation{Yunnan Observatories, Chinese Academy of Sciences, 396 Yangfangwang, Guandu District, Kunming, 650216, People's Republic of China}
\affiliation{Key Laboratory for the Structure and Evolution of Celestial Objects, Chinese Academy of Sciences, 396 Yangfangwang, Guandu District, Kunming, 650216, People's Republic of China}
\affiliation{Center for Astronomical Mega-Science, Chinese Academy of Sciences, 20A Datun Road, Chaoyang District, Beijing, 100012, People's Republic of China}
\affiliation{University of Chinese Academy of Sciences, Beijing 100049, People's Republic of China}


\author{Thomas Barclay} %
\affiliation{NASA Goddard Space Flight Center, 8800 Greenbelt Road, Greenbelt, MD 20771, USA}
\affiliation{University of Maryland, Baltimore County, 1000 Hilltop Cir, Baltimore, MD 21250, USA}

\author[0000-0002-6939-9211]{Tansu Daylan} %
\affiliation{Department of Physics, and Kavli Institute for Astrophysics and Space Research, Massachusetts Institute of Technology, 77 Massachusetts Ave., Cambridge, MA 02139, USA}
\affiliation{Kavli Fellow}

\author{Maximilian\ N.\ G\"unther} %
\affiliation{Department of Physics, and Kavli Institute for Astrophysics and Space Research, Massachusetts Institute of Technology, 77 Massachusetts Ave., Cambridge, MA 02139, USA}

\author[0000-0001-5941-2286]{J.\ J.\ Hermes} %
\affiliation{Department of Astronomy \& Institute for Astrophysical Research, Boston University, 725 Commonwealth Ave., Boston, MA 02215, USA}

\author[0000-0002-4715-9460]{Jon M.\ Jenkins} %
\affiliation{NASA Ames Research Center, Moffett Field, CA, 94035}

\author[0000-0001-9911-7388]{David W. Latham} %
\affiliation{Center for Astrophysics \textbar Harvard \& Smithsonian, 60 Garden St., Cambridge, MA 02138, USA}

\author[0000-0001-8172-0453]{Alan M.\ Levine} %
\affiliation{Department of Physics, and Kavli Institute for Astrophysics and Space Research, Massachusetts Institute of Technology, 77 Massachusetts Ave., Cambridge, MA 02139, USA}

\author{George R.\ Ricker} %
\affiliation{Department of Physics, and Kavli Institute for Astrophysics and Space Research, Massachusetts Institute of Technology, 77 Massachusetts Ave., Cambridge, MA 02139, USA}

\author[0000-0002-6892-6948]{Sara Seager} %
\affiliation{Department of Physics, and Kavli Institute for Astrophysics and Space Research, Massachusetts Institute of Technology, 77 Massachusetts Ave., Cambridge, MA 02139, USA}
\affiliation{Department of Earth, Atmospheric, and Planetary Sciences, Massachusetts Institute of Technology, 77 Massachusetts Ave., Cambridge, MA 02139, USA}
\affiliation{Department of Aeronautics and Astronautics, Massachusetts Institute of Technology, 77 Massachusetts Ave., Cambridge, MA 02139, USA}

\author[0000-0002-1836-3120]{Avi Shporer} %
\affiliation{Department of Physics and Kavli Institute for Astrophysics and Space Research, Massachusetts Institute of Technology, Cambridge, MA 02139, USA}

\author[0000-0002-6778-7552]{Joseph D.\ Twicken} %
\affiliation{SETI Institute, 189 Bernardo Ave., Suite 200, Mountain View, CA  94043, USA}
\affiliation{NASA Ames Research Center, Moffett Field, CA, 94035}

\author{Roland K.\ Vanderspek}
\affiliation{Department of Physics, and Kavli Institute for Astrophysics and Space Research, Massachusetts Institute of Technology, 77 Massachusetts Ave., Cambridge, MA 02139, USA}

\author[0000-0002-4265-047X]{Joshua N.\ Winn} %
\affiliation{Department of Astrophysical Sciences,
Princeton University, 4 Ivy Lane, Princeton, NJ 08544, USA}

\begin{abstract}
We present an analysis of the first 20-second cadence light curves obtained by the \tess\ space telescope during its extended mission. We find a precision improvement of 20-second data compared to 2-minute data for bright stars when binned to the same cadence ($\approx\,$10-25\% better for $T \lesssim 8$\,mag, reaching equal precision at $T \approx 13$\,mag), consistent with pre-flight expectations based on differences in cosmic ray mitigation algorithms. We present two results enabled by this improvement. First, we use 20-second data to detect oscillations in three solar analogs (\gpav, \ztuc\ and \pmen) and use asteroseismology to measure their radii, masses, densities and ages to $\approx$\,1\%, $\approx$\,3\%, $\approx$\,1\% and $\approx$\,20\% respectively, including systematic errors. Combining our asteroseismic ages with chromospheric activity measurements we find evidence that the spread in the activity-age relation is linked to stellar mass and thus convection-zone depth. Second, we combine 20-second data and published radial velocities to re-characterize \pmen\,c, which is now the closest transiting exoplanet for which detailed asteroseismology of the host star is possible. We show that \pmen\,c is located at the upper edge of the planet radius valley for its orbital period, confirming that it has likely retained a volatile atmosphere and that the ``asteroseismic radius valley' remains devoid of planets. Our analysis favors a low eccentricity for \pmen\,c ($<$\,0.1 at 68\% confidence), suggesting efficient tidal dissipation ($Q/k_{2,1} \lesssim 2400$) if it formed via high-eccentricity migration. Combined, these early results demonstrate the strong potential of \tess\ 20-second cadence data for stellar astrophysics and exoplanet science.
\end{abstract}


\section{Introduction}

Precise photometry of stars from space telescopes such as CoRoT \citep{baglin06} and \kep/K2 \citep{borucki08,howell14} has revolutionized stellar astrophysics and exoplanet science over the past two decades. An important characteristic of light curves provided by these missions is the sampling rate (observing cadence), which limits the timescales of astrophysical variability that can be measured. For example, oscillations of Sun-like stars, white dwarfs, and rapidly oscillating Ap stars occur on timescales of minutes \citep{aerts08,handler13}, requiring rapid sampling to unambiguously identify pulsation frequencies. While specialized techniques can be used to extract information above the Nyquist frequency \citep{murphy13,chaplin14}, shorter integration times also avoid amplitude attenuation caused by time-averaging and thus increase the signal-to-noise ratio. This is particularly important for Sun-like stars because they oscillate with low ($\approx$\,parts-per-million) amplitudes \citep{garcia19}. Fast sampling is also critical for resolving fast astrophysical transient phenomena such as stellar flares, which can occur on timescales of minutes \citep[e.g.][]{hawley14,davenport16}.

Rapid sampling is also important for transiting exoplanets. For example, resolving the transit ingress and egress duration in combination with a precise mean stellar density allows breaking of degeneracies between impact parameter and orbital eccentricity \citep{seager03,winn10,dawson12}. This is particularly powerful for characterizing eccentricities -- and thus dynamical formation histories -- of small (sub-Neptune sized) planets \citep{vaneylen15,xie16}, for which radial velocities often only provide weak eccentricity information. More broadly, impact parameter constraints enabled by well-sampled light curves result in more accurate planet-to-star radius ratios, which in the era of Gaia \citep{gaia} are sometimes the dominant factors in the error budgets of planet radii derived from transit photometry \citep{petigura20}. Finally, high cadence also enables more accurate characterizations of transit-timing variations, which provide mass and eccentricity constraints for small planets \citep[e.g.,][]{lissauer11,price14}.

Observing cadences for space telescopes are mostly set by onboard storage and bandwidth limitations, which in turn are tied to the spacecraft orbit. Early missions such as MOST \citep{walker03,matthews04}, BRITE \citep{weiss14b} and CoRoT provided sub-minute cadence photometry, but light curve durations and precisions were limited by Sun-synchronous orbits resulting in a small continuous viewing zone and significant straylight contamination \citep[e.g.][]{reegen06}. The \kep\ mission mitigated both effects through an Earth-trailing orbit, providing continuous, long-duration photometry with high precision. However, the onboard storage capacity limited the observing cadence to 30-minute sampling (long-cadence) for the $\approx$\,165,000 main target stars \citep{jenkins10}, with a subset of 512 stars per observing quarter observed with 1-minute sampling \citep[short-cadence,][]{gilliland10b}. \kep\ short-cadence observations  demonstrated the value of rapid sampling, for example by enabling the first systematic program that takes advantage of the synergy between asteroseismology and exoplanet science \citep{huber13,davies15, silva15,lundkvist16,kayhan19}, and remained a highly sought after resource for the duration of the \kep\ mission.

Thanks to its innovative orbit and large onboard storage, the NASA \tess\ mission \citep{ricker14} is currently providing unprecedented flexibility for space-based, high-precision and rapid photometry. During its two year prime mission, \tess\ provided 30-minute cadence observations for the entire field of view and observed 20,000 pre-selected targets at 2-minute cadence for each observing sector\footnote{TESS observed 16,000 targets in Sectors 1—3 after which the limit on the number of targets was increased to 20,000 stars.}. In its extended mission, \tess\ also produces light curves with 20-second cadence, in addition to 2-minute cadence targets and 10-minute cadence full-frame images, providing new opportunities for asteroseismology and characterizing transiting planets. Here, we present an analysis of 20-second light curves obtained during the first sectors of the \tess\ extended mission, including an asteroseismic analysis of nearby solar analogs and a re-characterization of \pmen\,c, the first transiting planet detected by \tess.

\section{Observations}

\subsection{Target Sample}

\begin{figure}
\begin{center}
\resizebox{\hsize}{!}{\includegraphics{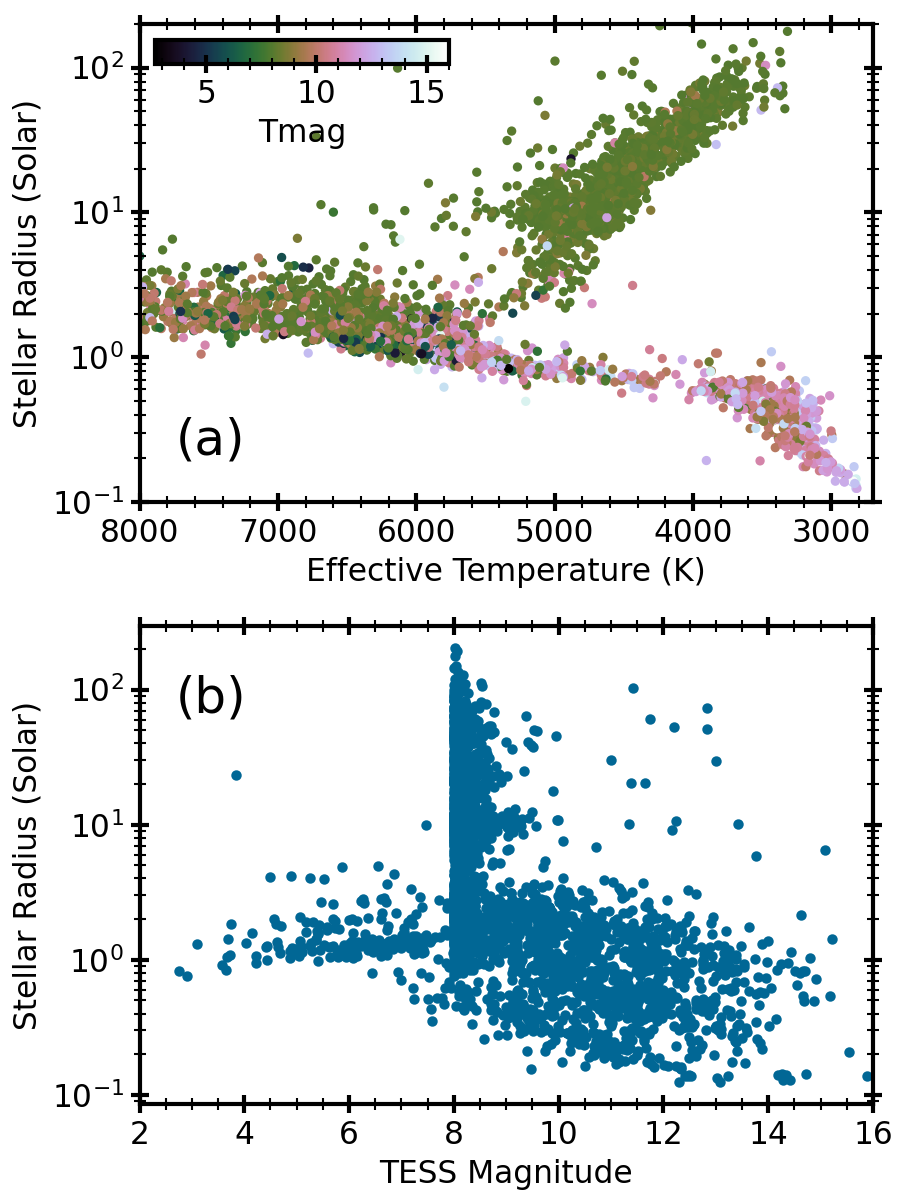}}
\caption{Panel (a): Stellar radius versus effective temperature of the $\approx$\,4400 unique stars with $\teff < 8000$\,K observed with TESS 20-second cadence observations during the first 10 sectors of the extended mission, color-coded by TESS magnitude $T$ (capped at $T<16$\,mag). Panel (b): Stellar radius versus TESS magnitude for the sample shown in panel (a). Compact stars (such as hot subdwarfs and white dwarfs) are excluded from both panels.}
\label{fig:hrd}
\end{center}
\end{figure}

\tess\ currently observes 1000 stars per sector at 20-second cadence during the extended mission, 600 of which are selected through the \tess\ Guest Investigator and Directors Discretionary programs. Figure \ref{fig:hrd} shows an H-R diagram and stellar radius versus \tess\ magnitude for the $\approx$\,4900 unique stars with $\teff < 8000$\,K observed during the first 10 sectors of the extended mission (July 4 2020 to April 2 2021), using effective temperatures and radii from the TESS Input Catalog \citep{stassun18,stassun19}. The \tess\ magnitude distribution in Figure \ref{fig:hrd}b shows a pile-up at $T\approx\,8$\,mag, which predominantly correspond to the 400 stars per sector observed for calibration purposes for the \tess\ Science Processing Operations Center \citep[SPOC,][]{jenkins16}. The calibration stars are requested to be bright but unsaturated, resulting in a tendency towards more evolved red giant stars. The main-sequence sample consists of a large number of optically faint M dwarfs, which are monitored to study stellar flares \citep{gunther20,feinstein20}, and a brighter sample of solar-type stars. Note that Figure \ref{fig:hrd} does not show compact stars (such as hot subdwarfs and white dwarfs), which are observed at 20-second cadence for asteroseismology \citep[e.g.][]{bell19,charpinet19} and to search for transiting planets \citep[e.g.][]{vanderburg20}.

\subsection{Photometric Performance}

\begin{figure*}
\begin{center}
\resizebox{\hsize}{!}{\includegraphics{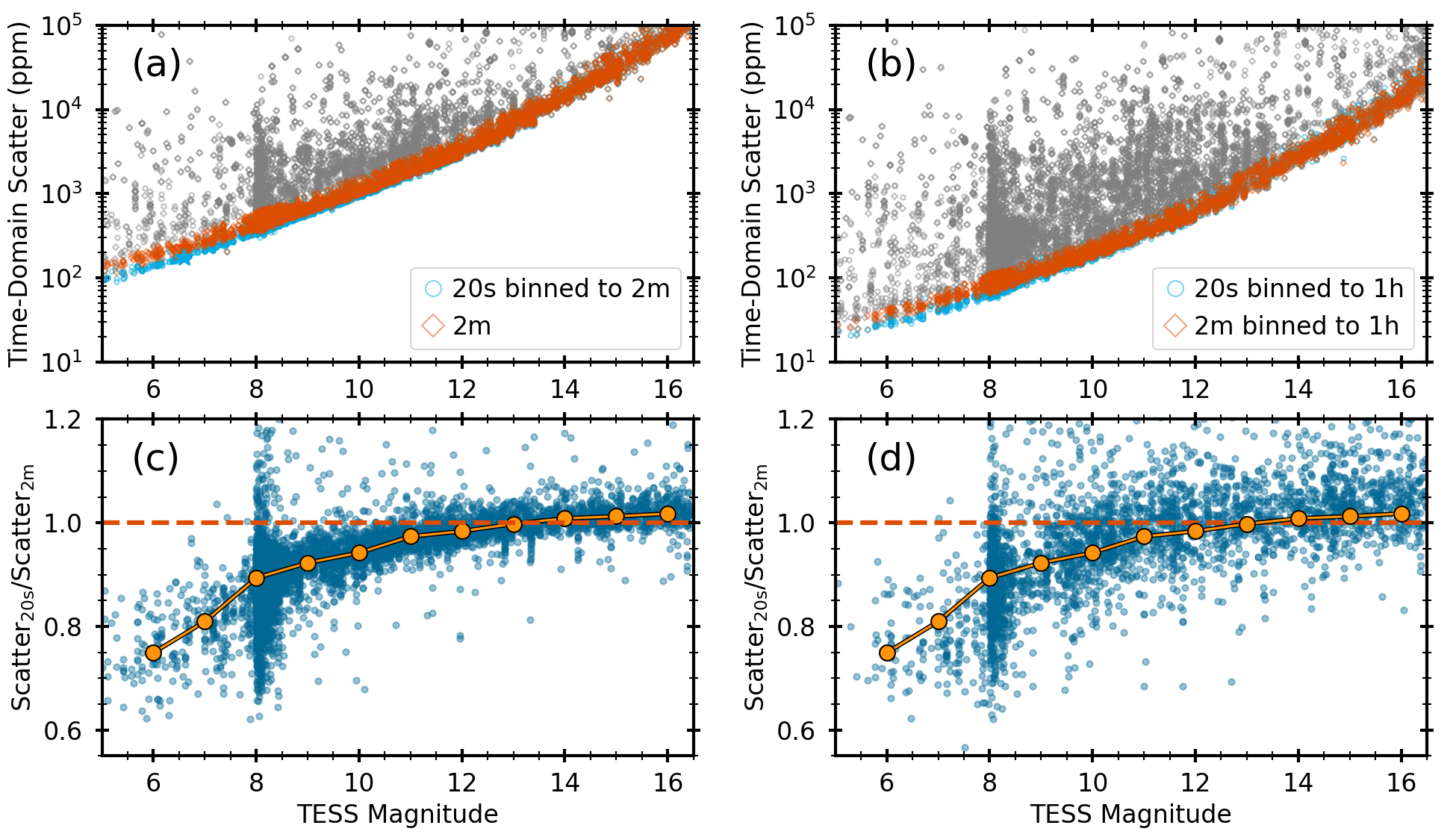}}
\caption{Panel (a): Time-domain scatter as a function of TESS magnitude for stars observed in 20-second cadence between Sectors 27 to 36. Symbols show 20-second light curves binned to 2-minute cadence (cyan circles) and original 2-minute light curves (red diamonds) for each sector. Grey points mark stars likely dominated by stellar variability (see text). Panel (b): Same as panel (a) but binning light curves to 1-hour cadence. Panel (c): Ratio of the time-domain scatter for the two datasets shown in panel (a), retaining only stars not dominated by stellar variability (i.e. each point in panel c is the ratio of a cyan circle and red diamond in panel a). The dashed line marks unity and orange circles show median bins. Panel (d): Same as panel (c) but for the binned light curves shown in panel (b).}
\label{fig:scatter}
\end{center}
\end{figure*}

To test the photometric precision, we downloaded all 20-second light curves obtained in Sectors 27-36 from the Mikulski Archive for Space Telescopes (MAST). We used the PDC-MAP light curves provided by the SPOC, which have been optimized to remove instrumental variability \citep{smith12,stumpe12,stumpe14}. We performed standard data-processing steps, retaining only data with quality flags set to zero. We then binned each light curve to 2-min cadence and 60-min cadence, high-pass filtered the data with a first order 0.5-day Savitzky-Golay filter \citep{savitzky64}, and calculated the standard deviation of the binned light curves (hereafter referred to as time-domain scatter) to provide a measure of the photometric precision on those timescales. We performed the same procedure using the original 2-minute light curves for the same stars, which are a standard SPOC data product and provide a benchmark for comparison to the new 20-second light curves. We calculated the photometric precision for each sector and each star to test the dependence of the noise properties on varying conditions between different sectors.

Figures \ref{fig:scatter}a and b show the measured time-domain scatter for each star and each sector as a function of TESS magnitude over 2-minute (left panels) and one hour timescales (right panels). As expected, the noise increases towards fainter magnitudes due to photon, sky and read noise. For each dataset, we identified stars dominated by stellar variability by scaling the TESS noise model from \citet{sullivan15} upward by 40\%, and marking all stars with a time-domain scatter above that level (grey points). 

\begin{figure}
\begin{center}
\resizebox{\hsize}{!}{\includegraphics{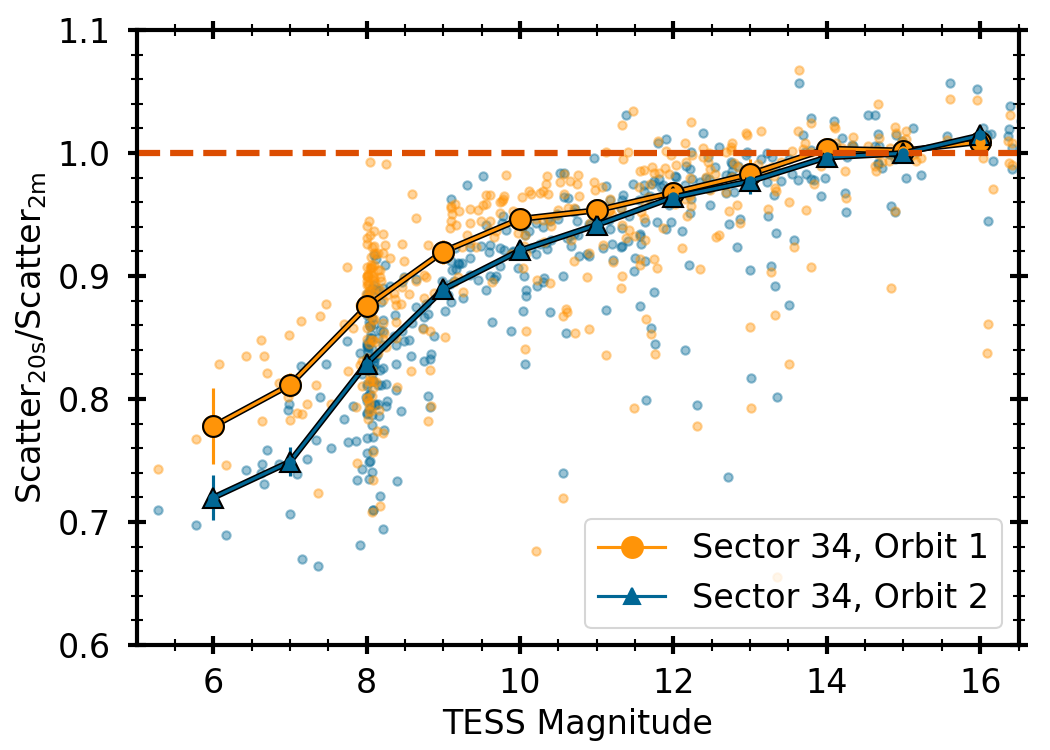}}
\caption{Ratio of the time-domain scatter for 20-second cadence data binned to 2-minute cadence and original 2-minute cadence light curves as a function of TESS magnitude for the first orbit (orange circles) and second orbit (blue triangles) of Sector 34 to test the impact of pointing jitter on the relative precision of the two light curve products (see text). Filled symbols show median bins in steps of 1 magnitude. The dashed line marks unity.}
\label{fig:scatters34}
\end{center}
\end{figure}

Figure \ref{fig:scatter} demonstrates that the 20-second light curves binned to 2-minute cadence show a strong magnitude-dependent improvement in precision compared to the original 2-minute cadence light curves. To illustrate this more clearly, Figure \ref{fig:scatter}c shows the ratio of the two measurements, again as a function of TESS magnitude. The average scatter for 20-second light curves is $\approx$\,25\% lower at $T=$\,6\,mag, $\approx$\,10\% lower at $T=$\,8\,mag and reaches parity with the 2-minute light curves around $T=$\,13\,mag. The same effect is seen for light curves binned to 1-hour cadence (Figure \ref{fig:scatter}d), but with larger scatter. Table \ref{tab:ratios} lists the median ratios in bins of one magnitude (orange circles in the bottom panels of Figure \ref{fig:scatter}) for each dataset, which may be used to approximate the precision of 20-second data relative to that of 2-minute data in the magnitude range $T=6-16$\,mag.


\begin{table}
\begin{center}
\caption{Noise ratios between 20-second and 2-minute data}
\begin{tabular}{c c c}
\tableline\tableline
\noalign{\smallskip}
TESS magnitude  & $[\sigma_{20s}/\sigma_{2m}]_{2m}$ & $[\sigma_{20s}/\sigma_{2m}]_{1h}$ \\
\hline  
6.0 & $ 0.749 \pm 0.007 $ & $ 0.766 \pm 0.011 $ \\
7.0 & $ 0.810 \pm 0.005 $ & $ 0.771 \pm 0.008 $ \\ 
8.0 & $ 0.894 \pm 0.001 $ & $ 0.877 \pm 0.003 $ \\
9.0 & $ 0.923 \pm 0.002 $ & $ 0.925 \pm 0.005 $ \\ 
10.0 & $ 0.942 \pm 0.002 $ & $ 0.955 \pm 0.003 $ \\
11.0 & $ 0.974 \pm 0.001 $ & $ 0.985 \pm 0.004 $ \\
12.0 & $ 0.984 \pm 0.001 $ & $ 0.996 \pm 0.004 $ \\
13.0 & $ 0.998 \pm 0.001 $ & $ 1.006 \pm 0.004 $ \\
14.0 & $ 1.009 \pm 0.001 $ & $ 1.012 \pm 0.004 $ \\
15.0 & $ 1.013 \pm 0.001 $ & $ 1.027 \pm 0.003 $ \\
16.0 & $ 1.018 \pm 0.002 $ & $ 1.032 \pm 0.003 $ \\
\hline
\end{tabular} 
\label{tab:ratios} 
\end{center}
\flushleft Note: Precision ratios apply for light curves retaining only quality flags set to zero. Keeping quality flags related to cosmic rays significantly degrades the 20-second data precision for bright stars (see text).
\end{table}

\begin{figure}
\begin{center}
\resizebox{\hsize}{!}{\includegraphics{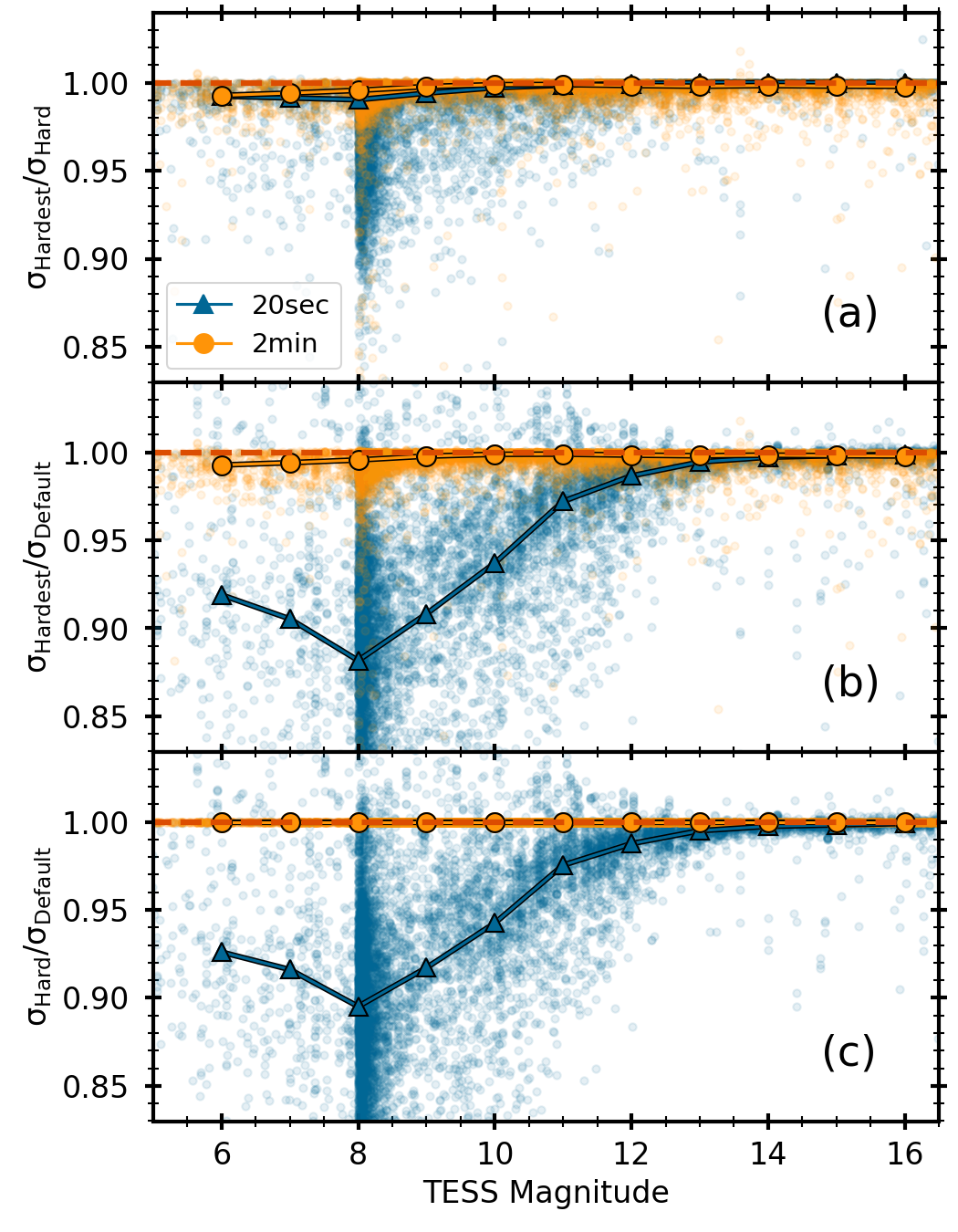}}
\caption{Ratio of the time-domain scatter for 20-second (blue triangles) and 2-minute (orange circles) data using three quality flag masks as defined in the Lightkurve package: ``Hardest'' (rejecting all data with non-zero quality), ``Hard'' (rejecting severe and cosmic ray flags) and ``Default'' (rejecting severe flags only). Points show individual sectors, filled symbols are median bins and the dashed line marks unity.}
\label{fig:qflags}
\end{center}
\end{figure}

The improvement shown in Figure 2 can partially be explained by the difference in cosmic-ray rejection algorithms applied to 20-second and 2-minute data. Specifically, 20-second data does not undergo onboard cosmic ray mitigation. Instead, cosmic ray mitigation is performed through post-processing by the SPOC, which identifies cadences affected by cosmic rays and attempts to correct their flux values. The onboard processing removes exposures with the highest and lowest flux for each stack of ten 2-second exposures, which leads to a 20\% reduction in effective exposure time for 2-minute data \citep{vanderspek18}. This shorter effective exposure time would correspond to a precision penalty of $\approx$10\% for 2-minute data if exposures were randomly rejected. Pre-flight simulations predicted a penalty closer to $\approx$3\% after taking into account that only exposures with the lowest and highest flux values are rejected (Z.\ Berta-Thompson, private communication). Since the improvement in Figure 2 is significantly larger than 3\%, this implies that sources in addition to photon noise must contribute to the distribution. 

\begin{figure*}
\begin{center}
\resizebox{\hsize}{!}{\includegraphics{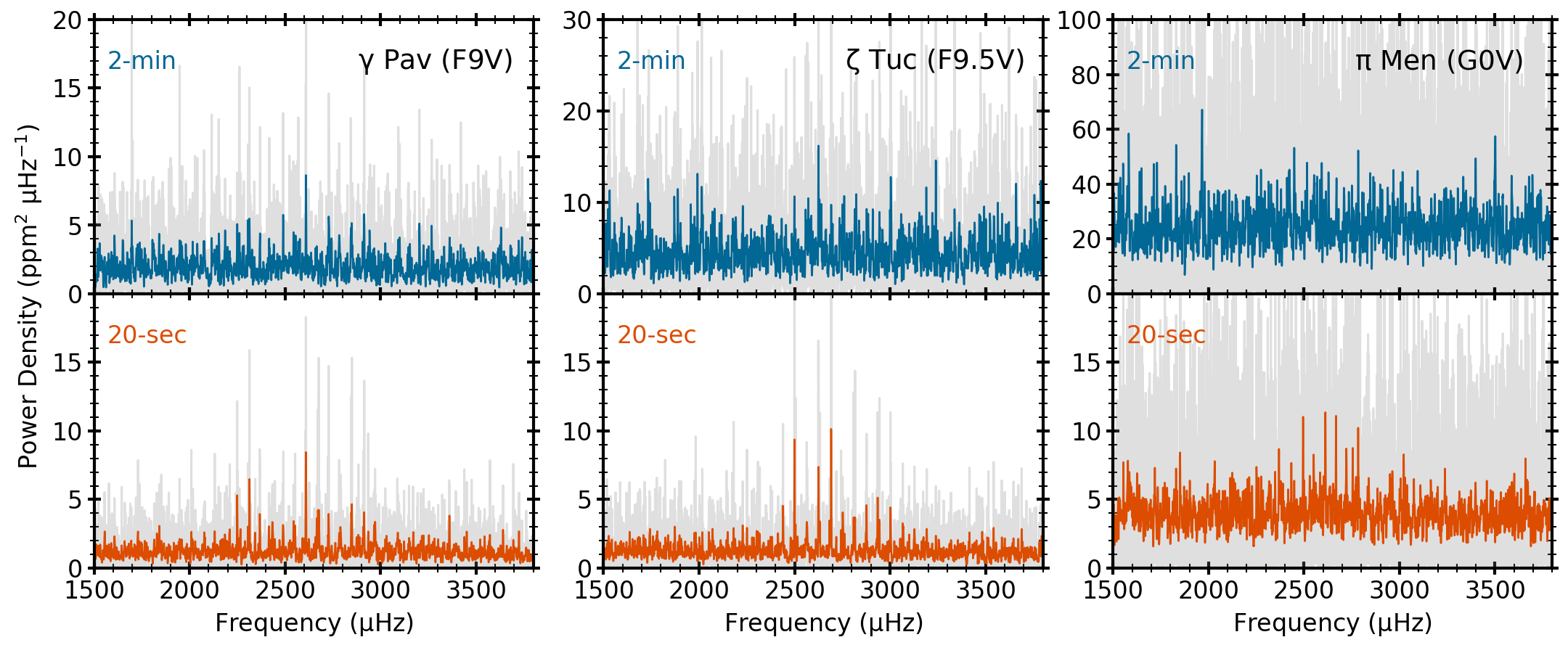}}
\caption{Power density spectra of the solar analogs $\gamma$\,Pav (left), $\zeta$\,Tuc (middle) and $\pi$\,Men (right) calculated using original 2-minute cadence data (top) and 20-second cadence data (bottom). Grey lines show original power spectra, colored lines show power spectra smoothed by 2\,\muHz. The S/N of the detection dramatically increases for the 20-second cadence data.}
\label{fig:compcadence}
\end{center}
\end{figure*}

One likely reason is pointing jitter, which for brighter stars should lead to larger changes in pixel values which are then preferentially removed during the onboard cosmic ray rejection for 2-minute data. To test this, we calculated time-domain scatter for the two halves of Sector 34, which had significantly different pointing performance. Figure \ref{fig:scatters34} confirms that the second orbit of Sector 34, which has larger pointing jitter, shows a stronger improvement of 20-second compared to 2-minute data, especially for the brightest stars. The improvement becomes negligible for stars fainter than $T\approx\,13$\,mag.

We also repeated the calculations using three different quality-flag masks as defined in the Lightkurve package (v2.0.10): ``Hardest'' (rejecting all data with non-zero quality flags, as done above), ``Hard'' (rejecting data with severe and cosmic ray flags only) and ``Default'' (rejecting data with severe flags only)\footnote{ \url{https://github.com/nasa/Lightkurve/blob/master/lightkurve/utils.py}. Note that the "Straylight" flag is currently not set in TESS 2-minute or 20-second data.}. Figure \ref{fig:qflags} compares the ratio of the time-domain scatter for the three mask combinations. For 2-minute data the mask choice has only a small impact ($\lesssim$\,1\% on average) and yields identical results when comparing the ``Hard'' and ``Default'' masks due to the on-board cosmic ray rejection. Removing data with cosmic-ray related quality flags yields significantly lower noise for 20-second data, especially for bright stars ($T<13$\,mag). The shape of the distribution is similar to the bottom panels of Figure \ref{fig:scatter}, demonstrating that removing cadences identified as cosmic rays during post-processing is important for the improved precision of 20-second data for bright stars. For faint stars ($T>13$\,mag) the choice of quality-flag mask has little influence on the precision for 20-second cadence data, which implies that the corrections applied to cadences affected by cosmic rays during post processing are more efficient for faint stars. We conclude that on average the best photometric precision for 20-second data is achieved when keeping only quality flags set to zero for bright stars ($T<13$\,mag).

The results presented here were predicted in pre-flight simulations, which showed that spacecraft jitter would lead to excess noise when applying the onboard cosmic ray mitigation for the brightest stars but would provide significant noise improvement for the larger number of faint stars (Z.\ Berta-Thompson, private communication). Additional effects that may impact the relative precision of 2-minute and 20-second light curves include the size of photometric apertures, which are calculated separately for each cadence. While a detailed investigation of these and other effects is left for future work, the confirmation of the pre-flight expectations presented here has significant ramifications for the allocation of 20-second cadence target slots, which are a scarce resource.  Specifically, for stars brighter than $T<13$\,mag (and especially for $T<8$\,mag) 20-second data provides  improved photometric precision irrespective of the timescale of astrophysical variability. Conversely, stars with $T>13$\,mag gain little from being observed in 20-second cadence unless the detection of astrophysical variability requires fast sampling (such as stellar flares or the detection of pulsations and transits for compact objects such as white dwarfs or subdwarfs).

\section{Asteroseismology}

\subsection{Oscillations in Solar Analogs}

To search for solar-like oscillations in the 20-second cadence sample, we analyzed all 84 solar-type stars observed as part of Cycle 3 Guest Investigator Program 3251\footnote{\url{https://heasarc.gsfc.nasa.gov/docs/tess/data/approved-programs/cycle3/G03251.txt}} (PI Huber). We performed the same  data processing steps described in the previous section and manually inspected the power spectra of each star. We only included data from Sectors 27 and 28 in this study.

We detected clear oscillations in three bright solar-like stars: \gpav\ (F9V, $V=4.2$\,mag), \ztuc\ (F9.5V, $V=4.2$\,mag) and \pmen\ (G0V, $V=5.7$\,mag). For \pmen, the SPOC light curve for Sector 27 showed scatter that is about a factor of two larger than expected. We therefore constructed a custom light curve from the target pixel files using the Lightkurve software package \citep{lightkurve}. We selected a larger aperture than had been used to construct the SPOC light curve, thereby capturing more of the flux from \pmen. The larger aperture was the single biggest factor in improving the quality of the light curve, and we confirmed that it captured all flux from saturated pixels. The light curve was extracted using simple aperture photometry. Then, to further correct for instrumental trends in the raw light curve, background pixels that were not within the target aperture were used to identify the four most significant trends via principal component analysis. The raw light curve was then detrended against these principal components, resulting in our corrected light curve. We created light curves for \pmen\ and \ztuc\ using this method for both 20-second and 2-minute cadence data. For \gpav\ we used regular SPOC PDC-MAP light curves.

Figure \ref{fig:compcadence} shows the power spectra of each star centered on the power excess due to solar-like oscillations. Note that we removed the transits of \pmen\,c from the light curve prior to our analysis.
The location of the power excess from oscillations predominantly depends on stellar surface gravity \citep{brown91}, and the observed excess at $\approx$\,2500\,\muHz\ for each star is consistent with predicted values from the \tess\ Asteroseismic Target List \citep[ATL,][]{schofield19}. For comparison, the top panels of Figure \ref{fig:compcadence} show power spectra calculated using the 2-minute cadence light curves of the same stars. We observe a strong improvement in S/N in all three stars, highlighting the benefit of the TESS 20-second light curve products for the study of solar-like oscillations in Sun-like stars. Indeed, this demonstrates that a few sectors of 2-minute data for $\pi$\,Men are insufficient for a detection of oscillations, which only become significant with 20-second cadence light curves. 

\begin{figure*}
\begin{center}
\resizebox{\hsize}{!}{\includegraphics{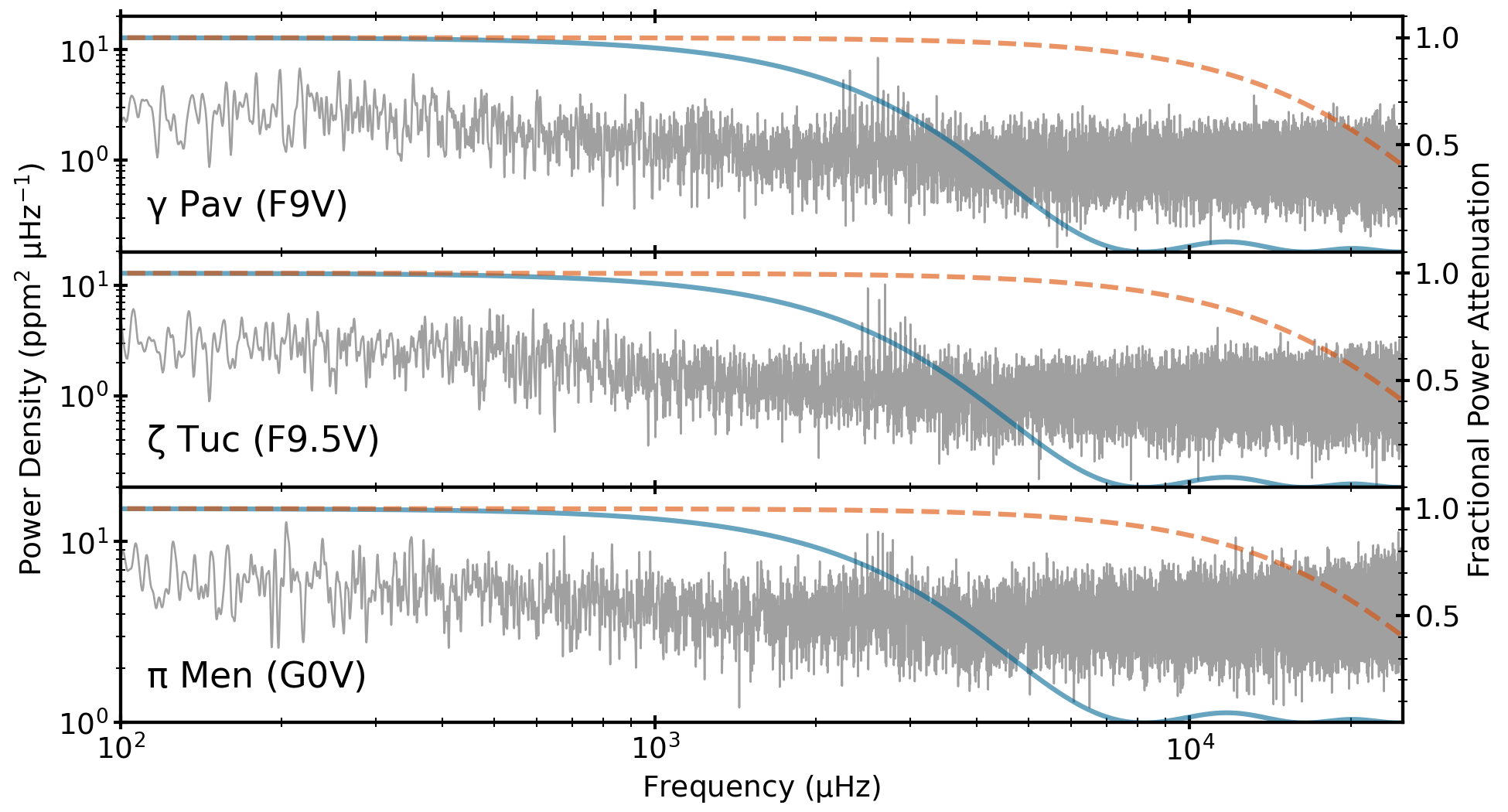}}
\caption{Power density spectra on a log-log scale using the 20-second cadence light curves for \gpav\ (top), \ztuc\ (middle) and \pmen\ (bottom). Each power spectrum was smoothed with a Gaussian with a FWHM of 1\,\muHz. Lines show the fractional power attenuation as given by Equation (2) on the right-hand y-axis for 2-minute cadence sampling (solid lines) and 20-second cadence sampling (dashed lines) in each panel.}
\label{fig:compcadence-log}
\end{center}
\end{figure*}

In addition to the lower time-domain noise, the S/N improvement in Figure \ref{fig:compcadence} can be attributed to the reduced amplitude attenuation enabled by the shorter integration times of 20-second cadence observations. The fractional amplitude attenuation caused by time-averaging of a signal with frequency $f$ is given by:

\begin{equation}
A = {\rm sinc}(\pi f t_{\rm exp}) \; ,
\end{equation}

where ${\rm sinc}(x) = (\sin x)/x$ and $t_{\rm exp}$ is the exposure time. For observations with no dead time between exposures such as those obtained by \tess\ and \kep, the exposure time is equal to the sampling time and thus the fractional attenuation in power can be written as:

\begin{equation}
P = {\rm sinc}^2\left({\pi \over 2} {f \over f_{\rm Nyq}} \right) \; ,
\label{equ:poweratt}
\end{equation}

where $f_{\rm Nyq} = 1/(2 \Delta t)$ is the Nyquist frequency for a timeseries with a constant sampling rate $\Delta t$. 

Figure \ref{fig:compcadence-log} shows the power spectra for the three stars on a log-log scale with lines showing the fractional power attenuation given by  Equation \ref{equ:poweratt}. Figure \ref{fig:compcadence-log} demonstrates that the longer sampling time of 2-minute data causes power attenuation of up to 30\% at frequencies corresponding to the power excess of Sun-like stars. In contrast, the rapid sampling for 20-second cadence alleviates power attenuation, demonstrating the importance of 20-second cadence data for the study of solar analogs using asteroseismology with TESS.

\subsection{Power Spectrum Analysis}

Several groups of coauthors used various analysis methods to extract global oscillation parameters \citep[e.g.][]{huber09,mosser09,mathur10b,mosser11c,benomar12,corsaro14,ref:lundkvist2015,stello17,campante18,nielsen21,pysyd}, many of which have been extensively tested on \kep, K2 and \tess\ data \citep[e.g.][]{hekker11,verner11,zinn20,stello21}.  In most of these analyses, the contributions due to granulation noise and stellar activity were modeled by a combination of Harvey-like functions \citep{harvey88} and a flat contribution due to photon noise.
The frequency of maximum power (\numax) was measured either by heavily smoothing the power spectrum or by fitting a Gaussian function to the power excess. We calculated final \numax\ values given in Table 5 as the median over a total of eleven different methods, with uncertainties calculated by adding in quadrature the standard deviations over all methods and the median formal uncertainty. The \numax\ measurement uncertainties range from $\approx$\,2-4\%.

To extract individual frequencies, different groups of coauthors applied either traditional iterative sine-wave fitting, i.e., pre-whitening \citep[e.g.][]{lenz05,kjeldsen05,bedding07} or Lorentzian mode-profile fitting \citep[e.g.][]{garcia09, handberg11,appourchaux12,mosser12,corsaro14, corsaro15,breton21}. For each star, we compared results and required at least two independent methods to return the same frequency within uncertainties. For the final list of frequencies we adopted values from one fitter who applied pre-whitening, with uncertainties derived by adding in quadrature the median formal uncertainty and the standard deviation of the extracted frequencies from all methods that identified a given mode. The frequency lists are given in Tables \ref{tab:gpav}, \ref{tab:ztuc} and \ref{tab:pmen}.

\begin{table}
\begin{center}
\caption{Extracted oscillation frequencies and mode identifications for \gpav.}
\vspace{0.1cm}
\begin{tabular}{c c c}        
$f(\muHz)$  & $\sigma_{f} (\muHz)$ & $l$ \\
\hline         
2249.47&0.42&1 \\
2305.83&0.86&2 \\ 
2313.73&0.34&0 \\
2367.90&0.58&1 \\
2425.37&1.79&2 \\
2433.42&0.70&0 \\
2490.23&0.75&1 \\
2545.02&0.89&2 \\
2552.39&0.68&0 \\ 
2609.23&0.48&1 \\
2666.19&0.99&2 \\
2672.37&1.04&0 \\
2728.45&0.68&1 \\
2783.03&1.53&2 \\
2790.16&1.10&0 \\
2849.84&0.84&1 \\
2906.72&2.46&2 \\
2912.60&1.14&0 \\
2971.50&1.32&1 \\
\hline         
\end{tabular}
\label{tab:gpav}
\end{center}
\end{table}

\begin{table}
\begin{center}
\caption{Same as Table \ref{tab:gpav} but for \ztuc.}
\vspace{0.1cm}
\begin{tabular}{c c c}        
$f(\muHz)$  & $\sigma_{f} (\muHz)$ & $l$ \\
\hline         
2439.55&0.46&0 \\
2499.60&0.32&1 \\
2558.06&0.99&2 \\
2565.84&1.04&0 \\
2625.23&0.41&1 \\
2682.98&1.12&2 \\
2691.73&0.49&0 \\
2752.10&0.81&1 \\
2809.75&0.61&2 \\
2816.76&0.46&0 \\
2876.80&0.44&1 \\
2935.26&0.64&2 \\
2944.52&0.72&0 \\
3002.66&0.51&1 \\
3069.40&1.04&0 \\
3127.60&0.73&1 \\
\hline         
\end{tabular}
\label{tab:ztuc}
\end{center}
\end{table}

\begin{table}
\begin{center}
\caption{Same as Table \ref{tab:gpav} but for \pmen.}
\vspace{0.1cm}
\begin{tabular}{c c c}        
$f(\muHz)$  & $\sigma_{f} (\muHz)$ & $l$ \\
\hline         
2368.76&1.42&2 \\
2433.31&0.86&1 \\
2494.91&0.82&0 \\
2550.41&0.78&1 \\
2603.63&1.56&2 \\
2611.63&1.21&0 \\
2667.03&0.58&1 \\
2721.50&1.11&2 \\
2783.09&0.80&1 \\
\hline         
\end{tabular}
\label{tab:pmen}
\end{center}
\end{table}

\begin{figure*}
\begin{center}
\resizebox{\hsize}{!}{\includegraphics{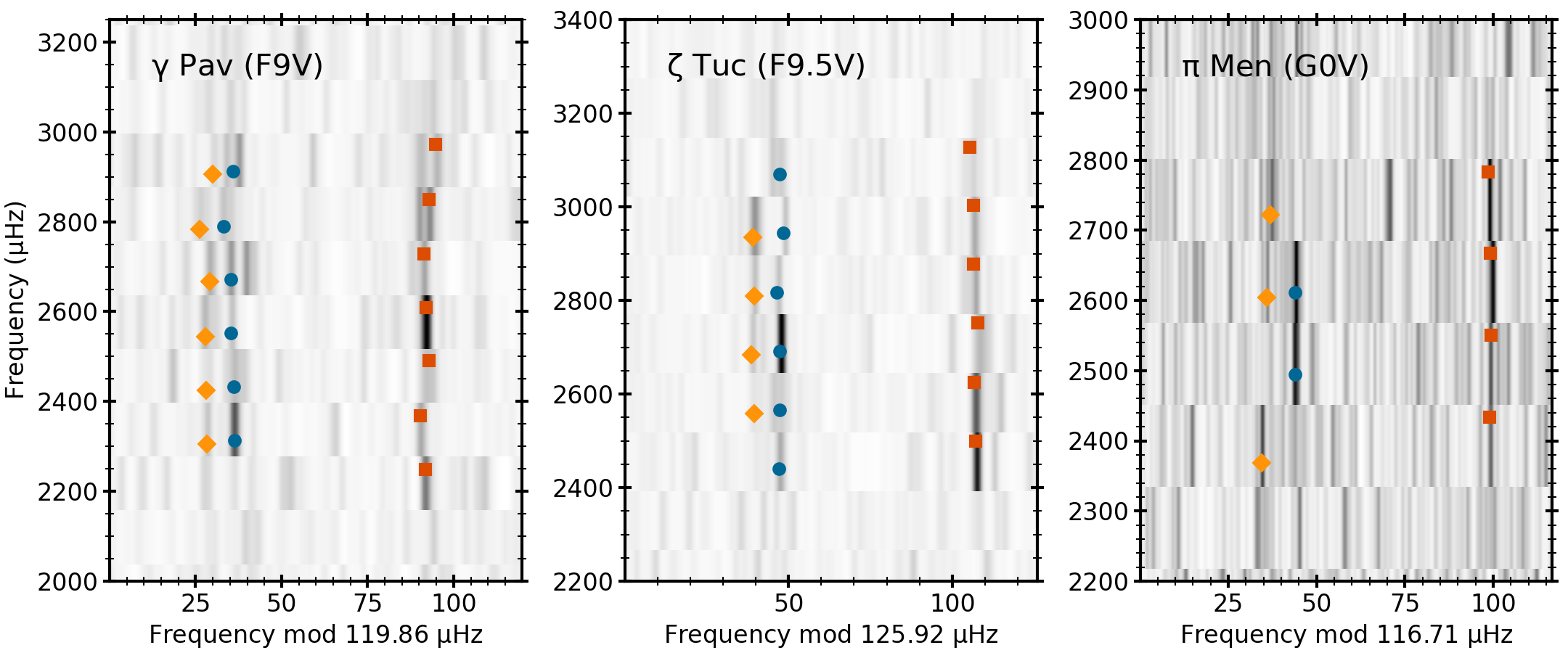}}
\caption{\'{E}chelle diagrams for \gpav\ (left), \ztuc\ (middle) and \pmen\ (right) using granulation-background corrected power spectra. Different symbols show measured radial modes (circles), dipole modes (squares) and quadrupole modes (diamonds). Error bars are smaller than symbol sizes in all cases.}
\label{fig:echelle}
\end{center}
\end{figure*}

To measure the large frequency separation, \dnu, we performed a weighted linear fit to all identified radial modes. Uncertainties were calculated by adding in quadrature the median formal uncertainty and the standard deviation for all estimates, yielding an average \dnu\ uncertainty of 0.8\% (Table \ref{tab:stellar}). Figure \ref{fig:echelle} shows the power spectra in \'{e}chelle format \citep{grec83} using these \dnu\ values, with extracted frequencies overlaid. As expected from Figure \ref{fig:compcadence} the frequency extraction was most successful for \gpav\ and \ztuc, yielding 6-7 dipole modes and strong constraints on the small frequency separation between modes with $l=0$ and 2, which is sensitive to the sound-speed gradient near the core and thus stellar age \citep{CD88}. The S/N for \pmen\ is lower due its fainter magnitude, but still allowed the extraction of several radial and non-radial modes. The offset of the $l=0$ ridge in each \'{e}chelle diagram, which is sensitive to the properties of the near-surface layers of the star \citep[e.g.][]{jcd14}, is consistent with expectations from \kep\ measurements for stars with similar $\dnu$ and $\teff$ \citep{white11}.

\subsection{Classical Constraints}

Due to the brightness of our stars, their atmospheric parameters such as effective temperature and metallicity have been extensively studied in the literature. We adopted \teff\ and \feh\ from \citet{aguilera18}, which were homogeneously derived from high-resolution spectroscopy. These values fall within $1\,\sigma$ of the median of \teff\ and \feh\ from 15 to 30 independent studies based on both photometry and spectroscopy. We adopted a 2\% systematic error in \teff, which accounts for uncertainties in the fundamental \teff\ scale based on the accuracy of angular diameters measured using optical long-baseline interferometry \citep{white18,tayar21}. This estimated uncertainty was added in quadrature to the formal 50\,K spectroscopic errors\footnote{All three stars have predicted angular diameters between $\approx$0.5-1\,mas, which can be resolved with current optical long-baseline interferometers. Measuring these angular diameters would be valuable to reduce the systematic uncertainties on \teff.}. We adopted a systematic uncertainty of 0.062\,dex in \feh\ to account for method-specific offsets \citep{torres12}. Note that \gpav\ is a metal-poor star with significant $\alpha$-element enhancement of $[\alpha/\rm{Fe}]=0.13\pm0.06$\,dex, calculated using individual abundances from
\citet[][]{bensby05}. Using the conversion by \citet{salaris93} yields $[\rm{M/H}]=-0.56\pm0.09$\,dex, which we adopted for model grids that do not specifically account for $\alpha$-element enhancement.

To calculate bolometric fluxes (\fbol), we fitted the spectral energy distribution of each target using broadband photometry following \citet{stassun16a}. Independent estimates were calculated from Tycho $V_T$ and $B_T$ photometry \citep{hog00}, combined with bolometric corrections from MIST isochrones \citep{choi16} as implemented in \texttt{isoclassify} \citep{huber17}. Interstellar extinction was found to be negligible in both methods, consistent with the short distances of all three targets. We also extracted \fbol\ estimates from the infrared flux method, as described in \citet{casagrande11}. Our final \fbol\ estimates were calculated as the median over all methods, with uncertainties calculated by adding the mean uncertainty and scatter over all methods in quadrature. The final \fbol\ uncertainties are 3 to 4\%, consistent with the expected systematic offsets \citep{zinn19,tayar21}. Finally, we combined \fbol\ values with \gaia\ EDR3 parallaxes \citep{lindegren21} to calculate luminosities, which provide an independent constraint for asteroseismic modeling. The results are summarized in Table \ref{tab:stellar}.

\subsection{Frequency Modeling}
\label{sec:modeling}

\begin{table*}
\begin{center}
\caption{Stellar Parameters}\label{tab:stellar}
\begin{tabular}{l c c c}
\tableline\tableline
\noalign{\smallskip}
 & $\gamma$\,Pav  & $\zeta$\,Tuc & $\pi$\,Men \\
\noalign{\smallskip}
\hline
\noalign{\smallskip}
Hipparcos ID & 1599 & 105858 & 26394 \\
HD Number & 203608 & 1581 & 39091 \\
TIC ID & 425935521 & 441462736 & 261136679 \\
$V_{T}$ Magnitude & 4.21 & 4.23 & 5.65 \\
\tess\ Magnitude & 3.67 & 3.72 & 5.11  \\
\hline
$\pi$ (mas) & \plxgpav & \plxztuc & \plxpmen  \\
\fbol ($10^{-7}$\,erg\,s$^{-1}$\,cm$^{-2}$) & \fbolgpav& \fbolztuc & \fbolpmen   \\
$L$ ($\lsun$) & \lumgpav& \lumztuc & \lumpmen   \\
\hline
\numax (\muHz) & \numaxgpav & \numaxztuc & \numaxpmen  \\
\dnu (\muHz) & \dnugpav & \dnuztuc & \dnupmen \\
\hline
\teff\, (K) & \teffgpav & \teffztuc & \teffpmen \\
\feh (dex) & \fehgpav & \fehztuc & \fehpmen \\
\hline
\mstar\ (\msun)& \massgpav & \massztuc & \masspmen \\
\rstar\ (\rsun)& \radgpav & \radztuc & \radpmen \\
\rhostar\ (gcc)& \dengpav & \denztuc & \denpmen \\
\logg\ (cgs) & \logggpav & \loggztuc & \loggpmen \\
Age (Gyr) & \agegpav & \ageztuc & \agepmen \\
\noalign{\smallskip}
\hline
\end{tabular}
\end{center}
\flushleft Notes: $V$ and \tess\ magnitudes are from the Tycho-2 catalog \citep{hog00} and TESS Input Catalog \citep{stassun18}, and parallaxes are from Gaia EDR3 \citep{lindegren21}. Effective temperatures and metallicities are from \citet{aguilera18}, with uncertainties calculated as described in the text. All other quantities are determined in this work. For modeling \gpav\ we adopted $[\rm{M/H}]=-0.56\pm0.09$\,dex based on $\alpha$-element abundances in \citet{bensby05} (see text).
\end{table*}

Different groups of coauthors used a number of approaches to model the observed oscillation frequencies, including different stellar evolution codes \citep[ASTEC, GARSTEC, MESA, and YREC,][]{jcd08,weiss08,paxton11,paxton13,paxton15,choi16,demarque08}, oscillation codes \citep[including ADIPLS and GYRE,][]{antia94,jcd08,gyre} and modeling methods \citep[including AIMS, AMP, ASTFIT, BeSSP, BASTA, PARAM and YB,][]{metcalfe09,stello09,basu10,gai11,creevey17,silva15,serenelli17,rodrigues14,rodrigues17,ong21,ball17,mosumgaard18, rendle19}. The adopted methods applied corrections for the surface effect \citep{kjeldsen08b,ball14}.
Model inputs included the spectroscopic temperature and metallicity, individual frequencies, \dnu, and luminosity. To investigate the effects of different input parameters, modelers were asked to provide solutions with and without taking into account the luminosity constraint from Gaia.

Overall, the modeling efforts yielded consistent results and we were able to provide adequate fits to the observed oscillation frequencies, as expected for stars with properties close to those of the Sun. The modeling results excluding and including the luminosity were consistent, demonstrating that there is no strong disagreement between the luminosity implied from asteroseismic constraints and from Gaia. To make use of the most observational constraints, we used the set of nine modeling solutions which used \teff, \feh, frequencies and the luminosity as input parameters. From this set of solutions, we adopted the self-consistent set of stellar parameters derived using MESA following \citet{ball17}, which showed the smallest difference to the median derived mass when averaged over all three stars. 

Table \ref{tab:stellar} lists our final stellar parameters for each star. For properties derived from asteroseismology (radius, mass, density, surface gravity and age) we quote random errors using the formal uncertainty of the adopted method following \citet{ball17} and systematic errors as  the standard deviation of the parameter over all methods. The results show that random and systematic errors have approximately equal contributions to the error budget, highlighting the importance of taking into account effects from different model grids. This is particularly pronounced for the mean stellar density, which formally can be measured with very high precision through the relation of the large separation to the sound speed integral \citep{ulrich86}. The average uncertainties (calculated by adding random and systematic errors in quadrature) are $\approx$\,1\% in radius, $\approx$\,3\% in mass, $\approx$\,1\% in density and $\approx$\,20\% in age, comparable to uncertainties from asteroseismology of Kepler stars \citep{silva17,celik21}. Systematic age uncertainties are largest for \gpav, consistent with larger differences in model predictions for metal-poor stars. 

\citet{mosser08} presented an asteroseismic analysis of \gpav\ based on five nights of radial velocity observations with HARPS. Our results show that the identification of even and odd degree modes was reversed in \citet{mosser08} due to the difficulty of ambiguously extracting frequencies from single-site ground-based data. Despite the different mode identification the derived mass is broadly compatible, but we measure a significantly younger age (5.9\,Gyr compared to 7.3\,Gyr). The younger age for \gpav\ derived here is consistent with asteroseismic red giant populations showing a flat age-metallicity relation (and thus mostly constant star formation history) for stars in the galactic disc \citep{casagrande16,silva18}.

\subsection{Activity-Age Relations}

Magnetic activity cycles are one of the most poorly understood aspects of stellar evolution, but play an important role for establishing empirical age indicators such as chromospheric activity \citep{mamajek08} and the spin-down of stars \citep[gyrochronology,][]{barnes03}. 
Stars with asteroseismic ages and characterized chromospheric activity cycles are critical ingredients for understanding and calibrating the interplay between rotation, age and activity. 
The measurement of activity cycles requires decades-long observations, which are typically only available for bright stars such as those included in the Mt.\ Wilson survey \citep{Baliunas95}. \tess\ has already demonstrated this powerful synergy for the solar analog $\alpha$\, Men \citep{chontos20} and the binary 94 Aqr \citep{metcalfe20}, thereby providing benchmarks for calibrating empirical age indicators.

\begin{figure}
\begin{center}
\includegraphics[width=\columnwidth,trim=0 0 40 30,clip]{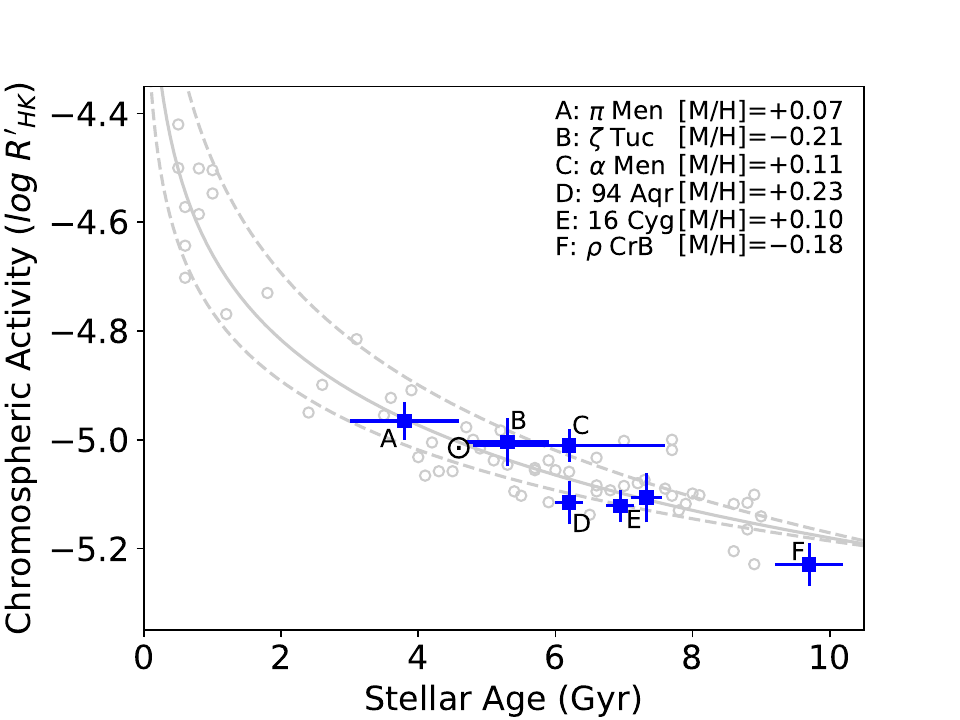}
\caption{Chromospheric activity versus stellar age for a sample of spectroscopic solar twins from \citet{lorenzo18} with ages determined from isochrone fitting (grey circles). The fitted relation is shown as a solid line, with uncertainties indicated by dotted lines. Overplotted are bright stars with asteroseismic ages from TESS and Kepler (blue symbols).}
\label{fig:act}
\end{center}
\end{figure}

The asteroseismic detections in bright solar analogs presented here provide additional benchmarks for calibrating activity-age relations. Figure~\ref{fig:act} shows the activity-age relation for a sample of spectroscopic solar twins from \citet{lorenzo18} with measured chromospheric activity from Ca\,{\sc ii} H\&K lines ($R'_{HK}$) and ages derived through isochrone fitting. We overplot several bright stars with asteroseismic ages \citep{chontos20,metcalfe20,creevey17,metcalfe21}, including the solar analogs in this paper, with $R'_{HK}$ placed on the same scale as \citet{lorenzo18} using the mean S-index values compiled by \citet{borosaikia18} and the $T_{\rm eff}$ values in Table~\ref{tab:stellar}. The resulting $R'_{HK}$ values were corrected for metallicity effects \citep[0.5$\times$\text{[M/H]} following][priv.~comm.]{saartesta12}. We omitted \gpav\ from the plot because its low metallicity falls outside of the calibration range. Error bars for the \citet{lorenzo18} sample were omitted because they do not take into account systematic errors from different model grids, as was done for the asteroseismic sample. 

Figure~\ref{fig:act} shows that the asteroseismic sample covers the critical regime at old ages ($\gtrsim$\,3\,Gyr) where the activity-age relation flattens. Interestingly, we observe that stars with similar ages and masses (such as the Sun and $\ztuc$) have similar $R'_{HK}$ values, while stars with similar ages but significantly different masses (such as $\alpha$\,Men\,A and 94\,Aqr\,Aa, with $0.94\msun$ and $1.22\msun$) show a significant spread in $R'_{HK}$.
This implies that the spread in the activity-age relation is probably linked to a spread in stellar mass and thus convection zone depth, analogous to the mass (or zero-age main-sequence temperature) dependence of gyrochronology relations \citep[e.g.][]{vansaders16}. Additional asteroseismic results from \tess\ 20-second data will be required to quantify such differential effects in activity-age relations. Additional extended mission observations in 20\,second cadence will also help to decrease the error bars on asteroseismic ages by enabling the detection of a larger number of oscillation frequencies and complement the existing database of active solar analogs already measured by \kep\ \citep{salabert16}.

\section{The $\pi$\,Men Planetary System}

\subsection{Asteroseismic Host Stars}

\begin{figure*}
\begin{center}
\resizebox{\hsize}{!}{\includegraphics{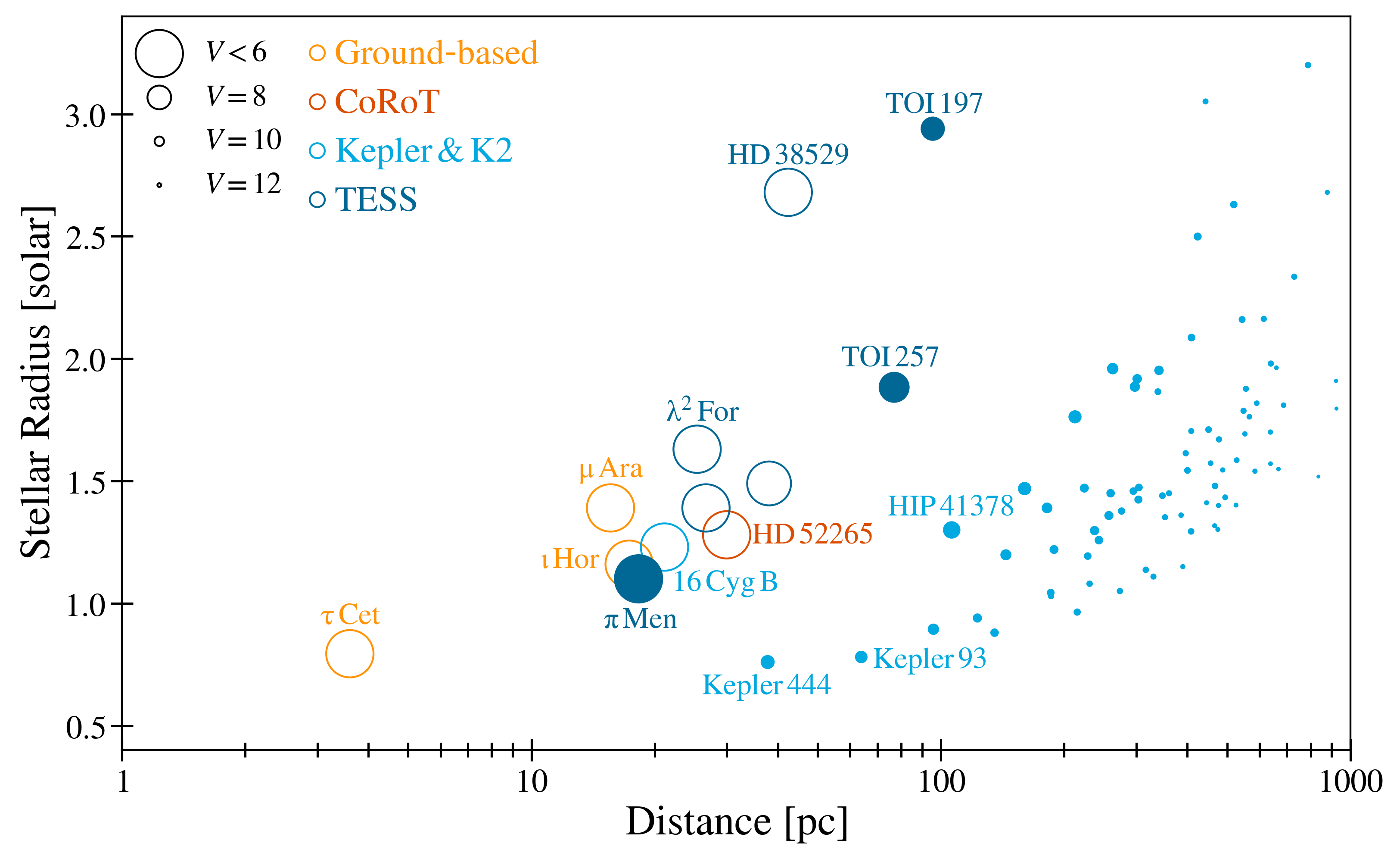}}
\caption{Stellar radius versus distance for the current population of asteroseismic host stars (excluding evolved stars with $R > 3.5 \rsun$) with confirmed \dnu\ measurements from ground-based radial velocity observations (orange), CoRoT (red), Kepler/K2 (light blue) and TESS (dark blue). Markers are sized by their visual magnitude and filled symbols show systems with known transiting planets. \pmen\ is now the closest and brightest star with a transiting planet for which detailed asteroseismic modeling is possible. \textbf{Label references (in order of proximity) --} $\tau$ Cet \citep{teixeira09}, $\mu$ Ara \citep{bouchy05}, $\iota$ Hor \citep{vauclair08}, $\pi$ Men (this work), 16 Cyg B \citep{metcalfe12}, $\lambda^2$ For \citep{nielsen20}, HD 52265 \citep{lebreton12,escobar12,lebreton14}, Kepler-444 \citep{campante15}, $\rm HD\,38529$ \citep{ball20}, Kepler-93 \citep{ballard14}, TOI-257 \citep{addison21}, TOI-197 \citep{huber19} and HIP 41378 \citep{vanderburg16,becker19,lund19}. } 
\label{fig:history}
\end{center}
\end{figure*}

\pmen\ joins the population of $\approx$\,110 exoplanet host stars which have been characterized using asteroseismology (Figure \ref{fig:history}). The majority of the sample comes from \kep\ \citep{huber13,lundkvist16}, which has led to important insights into demographics of small planet radii and eccentricities \citep{vaneylen15, vaneylen18, vaneylen19} and their dynamical formation histories through the measurement of asteroseismic spin-axis inclinations \citep{huber13b, chaplin13c, benomar14, lund14, campante16, kamiaka19,zhang21}. While the re-analyses of Kepler data and new data from the K2 Mission have added some detections \citep{vaneylen18b, chontos19, lund19}, the number of asteroseismic host stars has stagnated over the past few years.
Furthermore, similar to the general asteroseismic and host star sample, most \kep\ stars are faint and distant and thus difficult to characterize using ground-based observations. 

First results from \tess\ have already started to expand the population with detected oscillations in nearby exoplanet host stars, including newly discovered transiting exoplanets such as TOI-197 \citep{huber19} and TOI-257 \citep{addison21} and known exoplanet hosts \citep{campante19,nielsen20}. Ground-based radial velocity campaigns have yielded asteroseismic detections in some bright nearby exoplanet hosts including \pmen\ \citep{hodzic21}, but are generally limited to single-site observations causing ambiguities in the mode identification. The detection of oscillations presented here makes \pmen\ the closest and brightest star with a known transiting planet for which detailed asteroseismic modeling is possible, and highlights the strong potential of 20-second data to increase the asteroseismic host star sample.

\subsection{Transit and Radial Velocity Fit}

The 20-second cadence data provide an opportunity to re-characterize \pmen\,c, the first transiting planet discovered by \tess\ \citep{huang18}. In particular, the asteroseismic constraints on stellar radius and mean density, both measured with an \textit{accuracy} of $\approx$\,1\%, allows the opportunity to resolve degeneracies between eccentricity, impact parameter and transit duration. This degeneracy often limits the accuracy of derived planet radii \citep{petigura20} and can provide a constraint on the orbital eccentricities of small planets \citep{vaneylen15, vaneylen19}. In addition to the sub-Neptune sized \pmen\,c, the system includes a massive, non-transiting substellar companion on an eccentric orbit with an orbital period of $\approx$\,5.7\,years discovered using radial velocities \citep{jones02}.

We used \texttt{exoplanet} (v0.4.0) \citep{dfm21} to perform a joint fit of the TESS 20-second light curve and archival radial velocities spanning over $\approx$\,20 years from UCLES/AAT \citep{jones02}, HARPS \citep{huang18,gandolfi18}, CORALIE and ESPRESSO \citep{damasso20}. We follow \citet{damasso20} in splitting the HARPS, CORALIE and ESPRESSO datasets based on expected zeropoint offsets, calculating nightly bins, and parameterizing separate offsets and jitter terms ($\sigma$) for each of the eight radial velocity datasets. We also added a linear RV trend to the model to account for unknown outer companions. The transit model was parameterized with a photometric zeropoint offset, an extra photometric jitter term ($\sigma_{\rm TESS}$), conjunction times ($T_{0}$), orbital periods ($P$), impact parameters ($b$), quadratic limb darkening parameters ($u_{1},u_{2}$), eccentricity parameters ($\sqrt{e}\,\sin{\omega}$, $\sqrt{e}\,\cos{\omega}$), mean stellar density ($\rho_\star$), and radius ratio ($R_{p}/R_\star$). We included a Gaussian Process (GP) using a single simple harmonic oscillator kernel, consisting of a timescale ($\rho_{\rm GP}$), amplitude ($\sigma_{\rm GP}$), and a fixed quality factor $Q=1/\sqrt{2}$, to account for instrumental and stellar variability in the TESS light curve. For computational efficiency we only used 1 day chunks of the light curve centered on each transit. We used informative priors for the stellar mean density and radius based on the derived asteroseismic parameters (Table \ref{tab:stellar}) and wide Gaussian priors for the quadratic limb darkening parameters to account for uncertainties in model atmosphere predictions. The final model has 36 parameters, which were sampled using 4 chains with 1500 draws each and tested for convergence using the standard Gelman–Rubin statistic.  The priors and summary statistics for our joint transit and RV model are listed in Table \ref{tab:planet}. Our results agree well with previous analyses of the \pmen\ system \citep[e.g.][]{damasso20,gunther21}, but provide significantly improved eccentricity constraints for \pmen\,c (see Section 4.3).

\begin{figure*}
\begin{center}
\resizebox{\hsize}{!}{\includegraphics{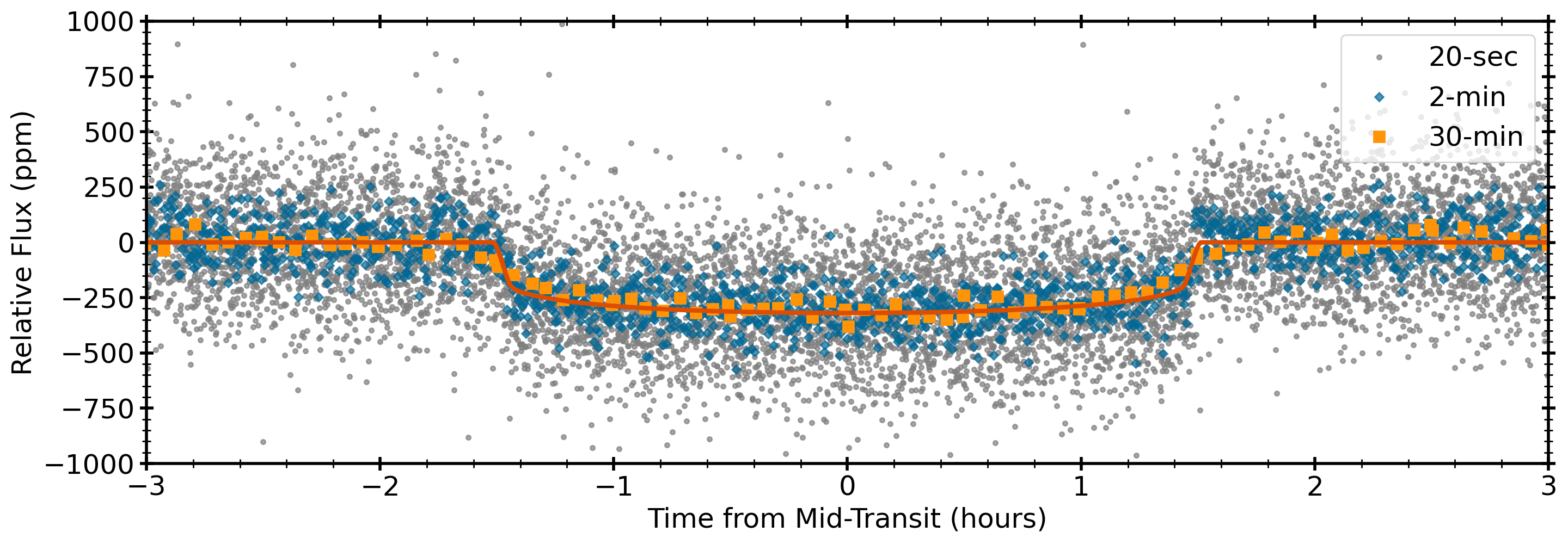}}
\caption{Phase folded transit light of \pmen\,c using Sector 27 and 28 20-second data after removal of a Gaussian process model to account for trends due to stellar and instrumental effects. Diamonds and squares show the original 20-second cadence data (grey circles) binned into 2-minute and 30-minute cadence. The best-fitting transit model is shown as the solid line. }
\label{fig:transitfit}
\end{center}
\end{figure*}

\begin{figure}
\begin{center}
\resizebox{\hsize}{!}{\includegraphics{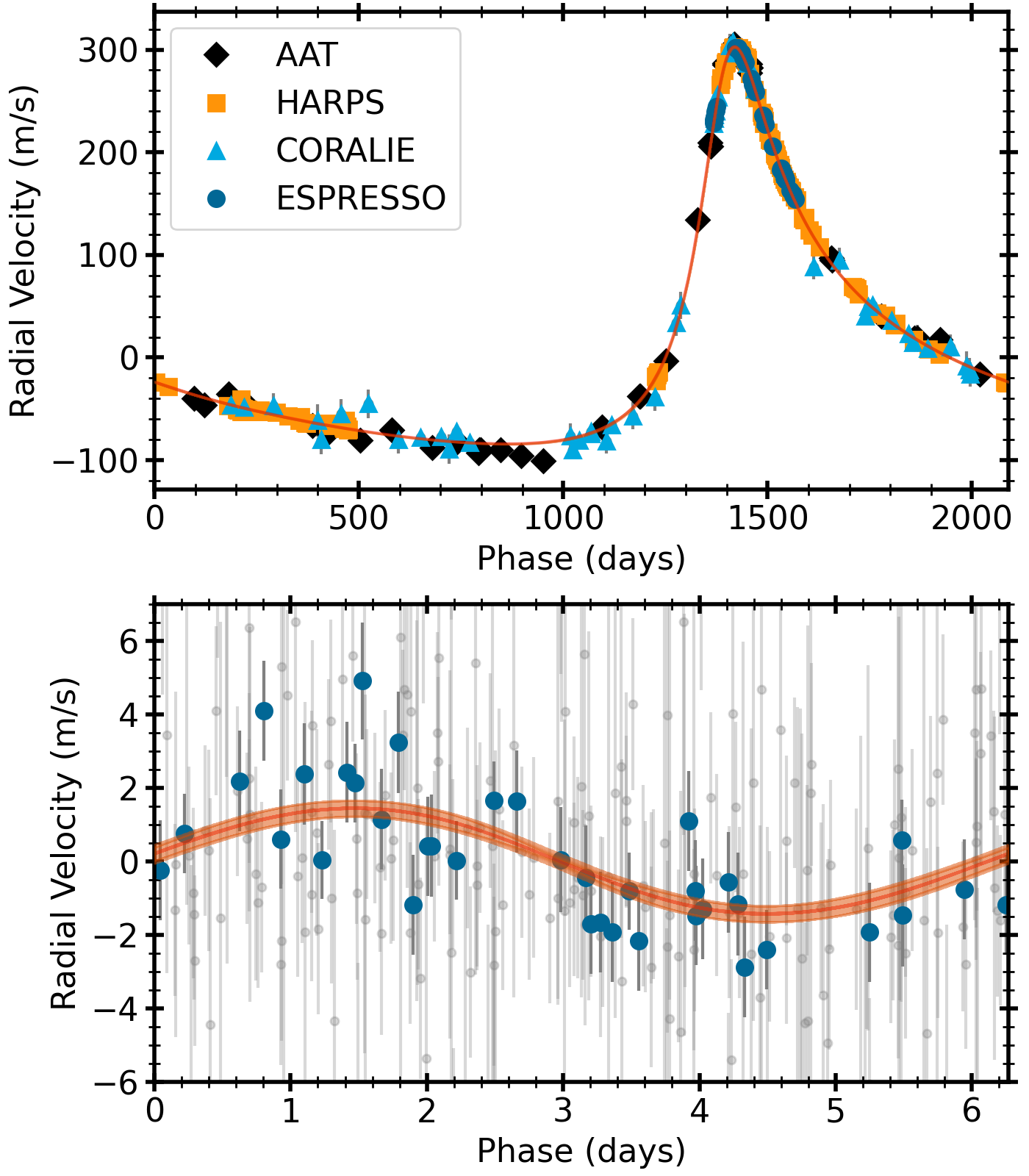}}
\caption{Radial velocity data phase-folded to the orbital periods of \pmen\,b (top panel) and \pmen\,c (bottom panel). The bottom panel shows all instruments except for ESPRESSO in grey. Each panel shows the RV data with contribution from the other planet removed.}
\label{fig:rvs}
\end{center}
\end{figure}

Figure \ref{fig:transitfit} shows the phase-folded transit light curve for \pmen\,c, and Figure \ref{fig:rvs} shows the radial velocity data for \pmen\,b and c with best-fitting models. Figure \ref{fig:transitfit} illustrates that the transits are well sampled in 20-second data, while 30-minute cadence would significantly smear out the ingress and egress durations. A detailed comparison of 20-second compared to 2-minute cadence for deriving transit parameters will be presented in a future study (Ho et al., in prep). As demonstrated by \citet{damasso20}, the combination of long-baseline and high-precision (in particular from ESPRESSO) for the available radial velocity dataset provides exquisite constraints on both planets (Figure \ref{fig:rvs}). We measure the radius and mass of \pmen\,c to $\approx$\,2\% and $\approx$\,13\%, which includes systematic errors on stellar parameters. 

\begin{table}
\begin{center}
\caption{\pmen\ Planet Parameters} \label{tab:planet}
\begin{tabular}{l c c}
\tableline\tableline
\noalign{\smallskip}
 Parameter & Prior & Value \\
\hline
\noalign{\smallskip}
$T_{0,b}$ \rm{(BTJD)}   & $\mathcal{N}(-466.8,1.0)$     & $-463.99^{+0.78}_{-0.76}$            \\
$T_{0,c}$ \rm{(BTJD)} & $\mathcal{N}(1519.8,0.1)$       & $1519.8016^{+0.0086}_{-0.0085}$        \\
$P_{b}$ (\rm{days}) & $\mathcal{N}(2090,10)$            & $2089.17^{+0.40}_{-0.40}$            \\
$P_{c}$ (\rm{days}) & $\mathcal{N}(6.27,0.01)$          & $6.26789^{+0.0001}_{-0.0001}$ \\
\rhostar & $\mathcal{N}(1.050,0.013)$                   & $1.052^{+0.013}_{-0.013}$ \\
$\sqrt{e_{b}}\,\cos{\omega_{b}}$ & $\mathcal{U}(-1,1)$  &  $0.7039^{+0.0015}_{-0.0014}$ \\
$\sqrt{e_{b}}\,\sin{\omega_{b}}$ & $\mathcal{U}(-1,1)$  &  $-0.3863^{+0.0036}_{-0.0033}$ \\
$\sqrt{e_{c}}\,\cos{\omega_{c}}$ & $\mathcal{U}(-1,1)$  &  $0.014^{+0.170}_{-0.172}$ \\
$\sqrt{e_{c}}\,\sin{\omega_{c}}$ & $\mathcal{U}(-1,1)$  &   $0.06^{+0.251}_{-0.251}$ \\
$b_{c}$ & $\mathcal{U}(0,1)$                            &  $0.591^{+0.056}_{-0.144}$ \\
$u_{1}$ & $\mathcal{N}(0.28,0.2)$                       &  $0.30^{+0.12}_{-0.12}$ \\
$u_{2}$ & $\mathcal{N}(0.28,0.2)$                       &  $0.22^{+0.15}_{-0.15}$ \\
$R_{p}/R_\star$ & $\mathcal{U}(0,1)$                    &  $0.01716^{+0.00024}_{-0.00030}$ \\
$K_{b}$ (m/s) & $\mathcal{N}(200,20)$                   & $194.55^{+2.83}_{-2.73}$ \\
$K_{c}$ (m/s) & $\mathcal{N}(1.5,1.5)$                  & $1.48^{+0.21}_{-0.20}$ \\
$\sigma_{\rm{GP}}$ (ppm)  &  IG$(3,500)$                & $51.1^{+10.9}_{-8.0}$ \\
$\rho_{\rm{GP}}$ (d)  &  $\log \mathcal{N}(5,10)$                       & $2.61^{+0.89}_{-0.64}$ \\
$\sigma_{\rm{TESS}}$ (ppm)  & $\log \mathcal{N}(0,10)$                  & $111.6^{+1.8}_{-1.8}$ \\
$\sigma_{\rm{AAT}}$ (m/s) & $\mathcal{U}(0.1,100)$                      & $4.46^{+1.16}_{-0.98}$ \\
$\sigma_{\rm{HARPS_{\rm{pre}}}}$ (m/s) & $\mathcal{U}(0.1,100)$         & $2.72^{+0.35}_{-0.30}$ \\
$\sigma_{\rm{HARPS_{\rm{post}}}}$ (m/s) & $\mathcal{U}(0.1,100)$        & $2.30^{+0.20}_{-0.17}$ \\
$\sigma_{\rm{CORALIE_{98}}}$ (m/s) & $\mathcal{U}(0.1,100)$             & $13.13^{+4.53}_{-3.03}$ \\
$\sigma_{\rm{CORALIE_{07}}}$ (m/s) & $\mathcal{U}(0.1,100)$             & $11.54^{+3.45}_{-2.42}$ \\
$\sigma_{\rm{CORALIE_{14}}}$ (m/s) & $\mathcal{U}(0.1,100)$             & $4.29^{+0.84}_{-0.75}$\\
$\sigma_{\rm{ESPRESSO_{\rm{pre}}}}$ (m/s) & $\mathcal{U}(0.1,100)$      & $1.04^{+0.40}_{-0.27}$ \\
$\sigma_{\rm{ESPRESSO_{\rm{post}}}}$ (m/s) & $\mathcal{U}(0.1,100)$     & $1.34^{+0.24}_{-0.20}$ \\
\hline
\multicolumn{3}{l}{\textbf{Derived parameters for \pmen\,b}} \\
$e_{b}$ & ---                & $0.6447^{+0.0011}_{-0.0012}$ \\
$\omega_{b}$ & ---           & $-28.75^{+0.27}_{-0.25}$\\
$a_{b}/R_\star$ & ---        & $625.6^{+2.5}_{-2.5}$ \\
$a_{b}$ (AU) & ---           & $3.315^{+0.031}_{-0.031}$ \\
$M\sin{i}_{b} (\mj)$ & ---   & $9.99^{+0.19}_{-0.19}$ \\
$M_{b} (\mj)$ & ---          & $13.07^{+1.16}_{-0.87}$ \\
\hline
\multicolumn{3}{l}{\textbf{Derived parameters for \pmen\,c}} \\
$e_{c}$ &  ---                   & $0.066^{+0.086}_{-0.047}$ \\
$\omega_{c}$ & ---               & $38.7^{+78.0}_{-135.2}$ \\
$a_{c}/R_\star$ & ---            & $12.977^{+0.052}_{-0.052}$ \\
$a_{c}$ (AU) & ---               & $0.06876^{+0.00065}_{-0.00065}$ \\
$i_{c}$ ($^{\text{o}}$) & ---    & $87.37^{+0.40}_{-0.11}$ \\
$R_{c} (\re)$ & ---              & $2.131^{+0.037}_{-0.042}$ \\
$M_{c} (\me)$ & ---              & $4.50^{+0.66}_{-0.63}$ \\
\noalign{\smallskip}
\hline
\end{tabular}
\end{center}
\flushleft Notes: See text for a description of all parameters. Nuisance parameters (photometric and RV instrument offsets, linear RV trend) are omitted from the table. $\mathcal{N}$, $\mathcal{U}$ and IG denote normal, uniform and inverse gamma distributions. The mass of \pmen\,b was calculated using $i=49.9\pm5.0^{\circ}$ \citep{derosa20}. 
\end{table}

\subsection{Dynamical Architecture}

Orbital eccentricities, inclinations and obliquities provide valuable information to constrain formation scenarios for close-in exoplanets. In particular, they help to distinguish dynamically ``hot'' formation pathways such as high-eccentricity migration triggered by planet-planet scattering \citep{chatterjee08,nagasawa08} or Kozai-Lidov cycles \citep{kozai62,lidov62,fabrycky07} from in-situ formation or migration in the protoplanetary disc \citep{cossou14}. While dynamical architectures have been extensively studied for hot Jupiters \citep[e.g.][]{winn10,albrecht12}, constraints for sub-Neptune sized planets are still relatively scarce, in particular for systems with known outer companions \citep{rubenzahl21}.

\pmen\ provides an excellent opportunity to study the dynamical formation pathway for a close-in sub-Neptune sized planet. The combination of Hipparcos and Gaia astrometry recently revealed that the orbit of \pmen\,b is  misaligned with \pmen\,c \citep{derosa20, xuan20}, while Rossiter-McLaughlin observations show a $24\pm4^{\circ}$ projected obliquity between the host star and \pmen\,c \citep{hodzic21}. Taken together these observations provide  evidence for a dynamically hot formation pathway for \pmen\,c. However, key dynamical properties such as the orbital eccentricity of \pmen\,c have so far been poorly constrained.

\begin{figure}
\begin{center}
\resizebox{\hsize}{!}{\includegraphics{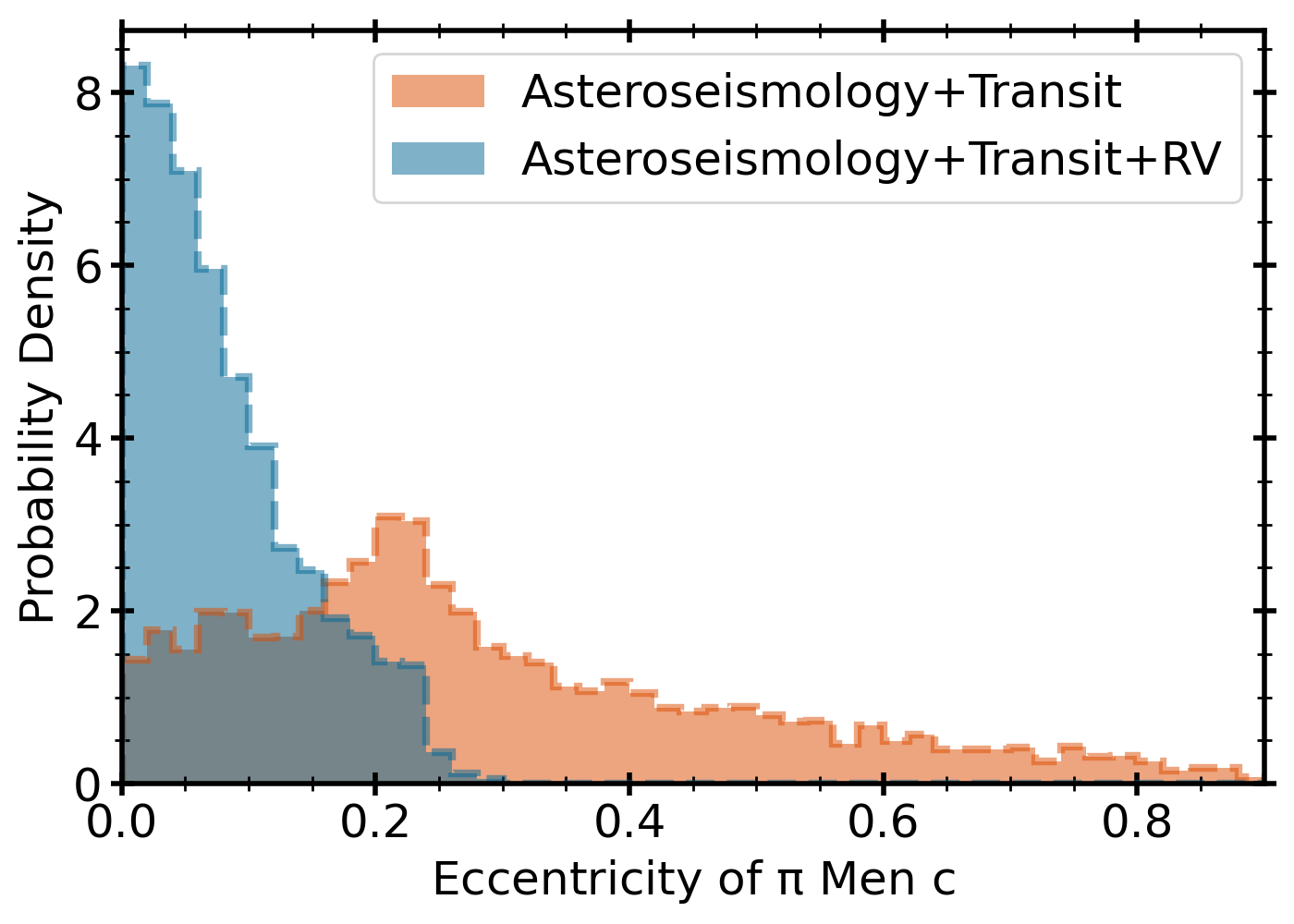}}
\caption{Marginalized posterior distribution on the orbital eccentricity of \pmen\,c based on the 20-second cadence transit fitting with asteroseismic stellar parameters alone (red) and including information from radial velocities (blue).}
\label{fig:ecc}
\end{center}
\end{figure}

Figure \ref{fig:ecc} shows the marginalized posterior distribution for the eccentricity of \pmen\,c derived from our fit to the 20-second photometry alone, and also using the joint transit and RV fit. Both are consistent with a low eccentricity for \pmen\,c and the joint fit places an upper limit of $<0.1$ (68\%), a factor of two tighter than previous constraints \citep{damasso20}. The posterior mode is consistent with a circular orbit, which implies that any initially high eccentricity caused by a dynamically hot formation has been damped over the $3.8\pm0.8$\,Gyr lifetime of the system. Tidal dissipation rates are highly uncertain, mainly owing to the unknown planetary tidal quality factor ($Q$) and tidal Love number ($k_{2,1}$), which quantify the strength of the energy dissipation of tides in the planet and the perturbation of the gravitational potential at its surface due to star-planet tidal interactions \citep{ogilvie14,mathis18}. Assuming that \pmen\,c has tidally circularized, we can place an upper limit on  $Q/k_{2,1}$ assuming equilibrium tides \citep{goldreich66,hut81,xuan20}:

\begin{equation}
\frac{Q}{k_{2,1}} < \frac{21 \pi}{2} \frac{M_{\star}}{M_{c}}\left(\frac{R_{c}}{a_{c}}\right)^{5} \frac{\tau}{P_{c}} \: ,
\end{equation}

\noindent
where $\tau$ is the age of the star and $P_{c}$ is the orbital period of \pmen\,c. Substituting values from Tables \ref{tab:stellar} and \ref{tab:planet} we arrive at $Q/k_{2,1} \lesssim 2400$. Assuming that \pmen\,c has a rocky core composed of iron and silicates in terrestrial proportions with a radius of $\approx$\,1.5\re\ and a mass of $\approx$\,4.5\,\me\ as inferred from planetary interior modeling \citep{garciamunoz21}, the ab-initio computations by \citet{tobie19} imply $k_{2,1} \approx 0.4$ for \pmen\,c \citep[compared to 0.3 for the Earth,][]{wahr81}. Using $Q/k_{2,1} \lesssim 2400$ leads to $Q \lesssim 970$, consistent with predictions that a $\approx$\,4.5\,\me\ planet with a terrestrial iron proportion should have $200 < Q < 1000$ depending on the considered viscosity \citep{tobie19}.

If \pmen\,c arrived at its present-day configuration through high-eccentricity migration, as suggested by the orbital misalignments, our data imply that it has completed tidal circularization. Assuming that the eccentricity excitation occurred through Kozai-Lidov cycles with \pmen\,b, this result would imply a present-day mutual inclination of $\pmen$\,c to $\pmen$\,b of $\sim$\,40$^{\circ}$ or $\sim$\,140$^{\circ}$ \citep{derosa20}. Furthermore, the circular orbit would suggest that \pmen\,c is not undergoing low-eccentricity migration, which has been suggested as a possible formation pathway for producing ultra-short period planets \citep{pu19}. Finally, the results imply that circular orbits for close-in sub-Neptune sized planet cannot be used to rule out dynamically hot formation scenarios. However, we note that dynamically ``cold'' formation pathways, such as disc migration or in-situ formation, can still explain the observed properties of \pmen\,c and thus cannot be completely excluded.

A key dynamical constraint for the \pmen\ system is the stellar spin-axis inclination, which would yield the full 3-D architecture of the system. While the current S/N is insufficient to reliably measure the spin-axis inclination using asteroseismology \citep{gizon03,ballot06,kamiaka18}, additional \tess\ 20-second observations (especially beyond the current extended mission, enabling $>$\,1 year coverage) may enable some constraints on this important parameter.

\subsection{Planet Radius Valley}

The dearth of planets with radii around 1.8\,\re\ in the Kepler sample \citep{fulton17} has sparked several efforts to investigate the origin and evolution of close-in sub-Neptune sized planets, including studies of small planets in the K2 sample \citep{hardegree20} and the dependence of the radius valley on both stellar mass \citep{fulton18,cloutier20,vaneylen21} and age \citep{berger20,david20,sandoval21}. A remarkable feature of the radius valley is that it is devoid of planets for a sample of well characterized stars and planets using asteroseismology \citep{vaneylen18}, suggesting that the dominant formation mechanism may be a relatively rapid process such as photoevaporation \citep{owen17}. While there is evidence that many ``gap planets'' in the general Kepler sample are linked to underestimated uncertainties in transit fits from long-cadence photometry \citep{petigura20}, recent studies have shown evidence for a transition of sub-Neptune to super-Earth sized planets on Gyr timescales \citep{berger20} and a shift of the radius gap with stellar age \citep{david20}, consistent with slower processes such as core-powered mass loss \citep{ginzburg18,gupta20}. If the primary mechanism for sculpting the radius gap operates on Gyr timescales, we should find examples of old planets with ages similar to \pmen\,c that are currently located in the gap.

\begin{figure}
\begin{center}
\resizebox{\hsize}{!}{\includegraphics{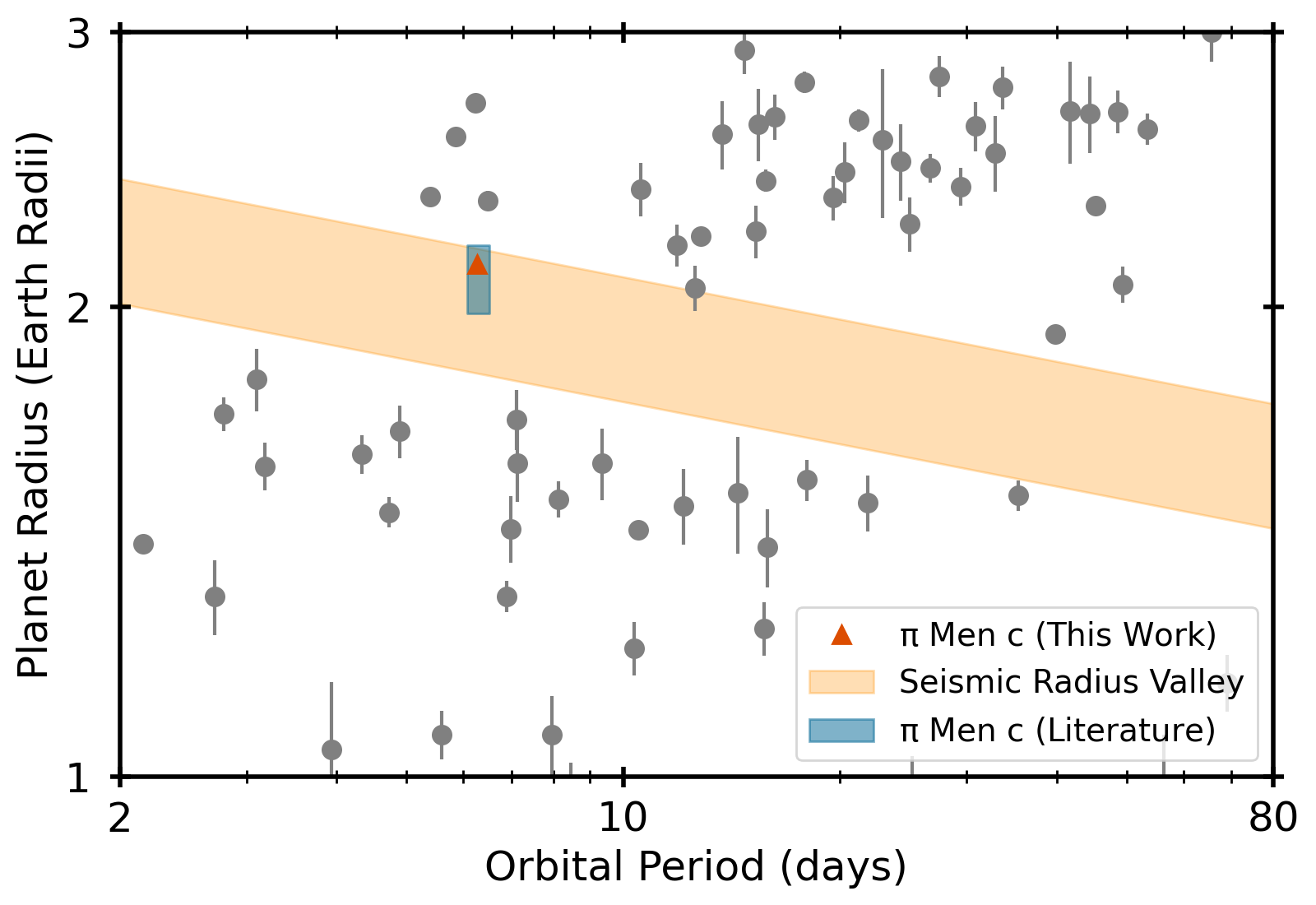}}
\caption{Planet radius versus orbital period for \kep\ exoplanets orbiting asteroseismic host stars from \citet{vaneylen18}. The shaded area shows the best-fit model for the radius gap as a function of orbital period from \citet{vaneylen18}. The rectangle covers published radii (including 1-$\sigma$ uncertainties) for \pmen\,c in the literature. The triangle marks the position of \pmen\,c from our analysis, with the error bar being smaller than the symbol size.}
\label{fig:valley}
\end{center}
\end{figure}

Figure \ref{fig:valley} compares our measured radius for \pmen\,c with the sample of small planets orbiting asteroseismic Kepler stars from \citet{vaneylen18}. Unlike some other studies, our analysis firmly places \pmen\,c  at the upper edge of the radius gap at its orbital period. 
This confirms previous results that \pmen\,c has probably held on to a significant volatile envelope even after 3.8\,Gyr and is consistent with recent HST transmission spectroscopy as well as planetary interior and long-term evolution modeling, which imply that \pmen\,c has a moderate to high molecular-weight atmosphere \citep{garciamunoz20,garciamunoz21}. 

The position of \pmen\,c at the upper edge of the radius gap also confirms the lack of genuine ``gap planets'' in the asteroseismic host star sample. Since the asteroseismic sample is biased towards older solar-type stars with ages ranging from $\approx$2-12\,Gyr and contains planets with well characterized radii, this suggests that the evolution of the radius valley may be restricted to $\lesssim$\,2\,Gyr.  However, we note that depending on the actual composition of its core and envelope it is also possible that the \pmen\,c will eventually evolve through the radius gap.

\section{Conclusions}

We have presented an analysis of the new 20-second cadence light curves provided by the \tess\ space telescope in its first extended mission. Our main conclusions are as follows:

\begin{itemize}
    \item \tess\ 20-second light curves show $\approx$\,10-25\% better precision than 2-minute light curves for bright stars with $T \lesssim 8$\,mag, reaching equal precision at $T \approx 13$\,mag. The improved precision is consistent with pre-flight expectations and can partially be explained by the increased effective exposure time for 20-second data due to the lack of on-board cosmic ray rejection and the decreased efficiency of the on-board cosmic ray rejection for 2-minute data in bright stars due to spacecraft pointing jitter.
    The results imply that \tess\ 20-second data are particularly valuable for bright stars since they yield improved photometric precision irrespective of the timescale of astrophysical variability.
    
    \item We use 20-second data to detect oscillations in three bright solar analogs observed in Sectors 27 and 28: \gpav\ (F9V, $V=4.2$), \ztuc\ (F9.5V, $V=4.2$) and \pmen\ (G0V, $V=5.7$). We used asteroseismology to measure their radii to $\approx\,1$\%, masses to $\approx\,3$\%, densities to $\approx\,1$\% and ages to $\approx\,20$\%, including systematic errors estimated by using different model grids and methods. We combine our asteroseismic ages with chromospheric activity measurements and find evidence that the spread in the activity-age relation is linked to stellar mass and thus convection zone depth.
    
    \item We combined asteroseismic stellar parameters, 20-second transit data and published radial velocities to re-characterize \pmen\,c, which is now the closest transiting exoplanet for which detailed asteroseismic characterization of the host star is possible. We measured the radius ($R=2.13\pm0.04$\re) and mass ($M=4.5\pm0.6$\me) to 2\% and 13\%, respectively. Our results show that \pmen\,c sits at the upper edge of the planet radius valley, suggesting that it has probably held on to a volatile atmosphere. The planet radius valley, considering only exoplanets orbiting $\approx$\,2-12\,Gyr old solar-type stars for which the precise asteroseismic characterization has been possible, remains devoid of planets.
    
    \item Our analysis provides strong evidence for a circular orbit for \pmen\,c ($e<0.1$ at 68\% confidence, with a mode consistent with zero). If \pmen\,c arrived at its present orbit through high-eccentricity migration, as suggested by its misalignment with the outer substellar companion \pmen\,b and the host star, our results imply that it has efficiently completed tidal circularization ($Q/k_{2,1} \lesssim 2400$ for the asteroseismic system age of $3.8\pm0.8$\,Gyr) and that circular orbits for close-in sub-Neptune sized planets alone cannot be used to rule out dynamically hot formation scenarios.
\end{itemize}

Continued 20-second cadence observations in the TESS extended mission would yield the opportunity for an asteroseismic catalog of bright solar analogs, which could be used to calibrate activity-age-rotation relationships for stars that have long-term activity monitoring. Additionally, fast sampling will continue to enable the opportunity to constrain orbital eccentricities for small planets from transit durations and expand the sample of host stars for which asteroseismic characterization is possible. The early results presented here demonstrate the strong potential of \tess\ 20-second data for stellar astrophysics and exoplanet science in the first extended mission and beyond.

Data and scripts to reproduce results and figures are available on GitHub\footnote{\url{https://github.com/danxhuber/tess20sec}} and version 1.0.0 is archived in Zenodo \citep{tess20sec}.

\vspace{0.5cm}
\noindent
We thank the entire TESS team for making 20-second cadence observations possible. We also thank Zach Berta-Thompson and John Doty for helpful discussions on \tess\ cosmic ray rejection algorithms and pre-flight simulations, and Kosmas Gazeas for helpful comments provided through the TASC review process.

D.H.~acknowledges support from the Alfred P. Sloan Foundation, the National Aeronautics and Space Administration (80NSSC21K0652), and the National Science Foundation (AST-1717000).
T.S.M.~acknowledges support from NASA grant 80NSSC20K0458. Computational time at the Texas Advanced Computing Center was provided through XSEDE allocation TG-AST090107. 
A.C.~acknowledges support from the National Science Foundation through the Graduate Research Fellowship Program (DGE 1842402).
W.H.B.~performed computations using the University of Birmingham's BlueBEAR High Performance Computing service.
T.R.B.~acknowledges support from the Australian Research Council through Discovery Project DP210103119.
Funding for the Stellar Astrophysics Centre is provided by The Danish National Research Foundation (Grant DNRF106).
M.S.C.\ and M.D.\ acknowledge the support by FCT/MCTES through the research grants UIDB/04434/2020, UIDP/04434/2020 and PTDC/FIS-AST/30389/2017, and by FEDER - Fundo Europeu de Desenvolvimento Regional through COMPETE2020 - Programa Operacional Competitividade e Internacionalização (grant: POCI-01-0145-FEDER-030389).
T.L.C.~is supported by Funda\c c\~ao para a Ci\^encia e a Tecnologia (FCT) in the form of a work contract (CEECIND/00476/2018).
M.S.C.~is supported by national funds through FCT in the form of a work contract.
HK and EP acknowledge the grant from the European Social Fund via the Lithuanian Science Council (LMTLT) Grant No. 09.3.3-LMT-K-712-01-0103.
R.A.G.~and S.N.B.~acknowledge the support received from the CNES with the PLATO and GOLF grants.
B.N. acknowledges postdoctoral funding from the Alexander von Humboldt Foundation and ``Branco Weiss fellowship Science in Society'' through the SEISMIC stellar interior physics group.
S.M.~acknowledges support by the Spanish Ministry of Science and Innovation with the Ramon y Cajal fellowship number RYC-2015-17697 and the grant number PID2019-107187GB-I00. 
T.W.\ acknowledges support from the B-type Strategic Priority Program of the Chinese Academy of Sciences (Grant No. XDB41000000), from the NSFC of China (Grant Nos. 11773064, 11873084, and 11521303), from the Youth Innovation Promotion Association of Chinese Academy of Sciences, and from the Ten Thousand Talents Program of Yunnan for Top-notch Young Talents. T.W.\ also gratefully acknowledges the computing time granted by the Yunnan Observatories and provided by the facilities at the Yunnan Observatories Supercomputing Platform.
T.D.\ acknowledges support from MIT's Kavli Institute as a Kavli postdoctoral fellow.

Funding for the TESS mission is provided by NASA's Science Mission directorate. 
Resources supporting this work were provided by the NASA High-End Computing (HEC) Program through the NASA Advanced Supercomputing (NAS) Division at Ames Research Center for the production of the SPOC data products.
This paper includes data collected by the TESS mission, which are publicly available from the Mikulski Archive for Space Telescopes (MAST). 

\software{This research made use of \textsf{exoplanet} \citep{dfm21} and its
dependencies \citep{exoplanet:agol20, exoplanet:astropy13, exoplanet:astropy18, astropy18,
exoplanet:exoplanet, exoplanet:luger18, exoplanet:pymc3, exoplanet:theano}, DIAMONDS \citep{corsaro14}, echelle \citep{echelle}, exoplanet \citep{exoplanet}, isoclassify \citep{huber17,berger20}, Lightkurve \citep{lightkurve}, Matplotlib \citep{matplotlib}, numpy \citep{numpy} and scipy \citep{scipy}.}

\bibliography{references}{}
\bibliographystyle{aasjournal}

\suppressAffiliationsfalse
\allauthors

\end{document}